\documentclass[useAMS,usenatbib]{mn2e}

\usepackage[pdftex]{graphicx}

\newcommand{\halp}{H$\alpha$}

\usepackage{relsize}

\newcommand{\hi}{%
  \relax
  \ifmmode
    \textrm{\textsc{HI}}
  \else
    \textsc{H{\smaller}I}
  \fi
}

\newcommand{\oh}{OH }
\newcommand{\co}{CO }

\newcommand{\simgt}{\lower.5ex\hbox{$\; \buildrel > \over \sim \;$}}
\newcommand{\simlt}{\lower.5ex\hbox{$\; \buildrel < \over \sim \;$}}

\title[The Shape of Dark Matter Haloes I.]{The Shape of Dark Matter Haloes\\
I. \hi  Observations of Edge-on Galaxies}
\author[S. P. C. Peters et al.]{S. P. C. Peters$^{1}$,
P. C. van der Kruit$^{1}$\thanks{For more information, please contact P.C. van der Kruit at vdkruit@astro.rug.nl.}, 
R. J. Allen$^{2}$ and K. C. Freeman$^{3}$\\
$^{1}$Kapteyn Astronomical Institute, University of Groningen, P.O.Box 800, 9700AV Groningen, the Netherlands\\
$^{2}$Space Telescope Science Institute, 3700 San Martin Drive, Baltimore, MD 21218, USA\\
$^{3}$Research School of Astronomy and Astrophysics The Australian National University, Cotter Road Weston Creek, ACT 2611,\\
Australia}

\begin{document}

\date{Accepted 2015 month xx. Received 2015 Month xx; in original form 2015 Month xx}
\pagerange{\pageref{firstpage}--\pageref{lastpage}} \pubyear{2015}

\maketitle

\label{firstpage}

\begin{abstract}
We present neutral hydrogen observations for a sample of eight nearby, 
late-type, edge-on galaxies.
All of the galaxies have been well resolved in the radial direction, while 
six have also been well resolved in the vertical direction.
We find that each of the galaxies has approximately the same maximum surface 
brightness temperature throughout its disc.
We argue that self-absorption is the main cause of this phenomenon and that 
subsequent decompositions will require a treatment of this.
\end{abstract}


\section{Introduction to the series}

Rotation curve of galaxies remain flat at large radii, implying that the 
amount of mass-to-light increases drastically with radius from the centre.
This first indication, in the early seventies by 
\citet{Freeman1970A}, was that in two galaxies the observed \hi-rotation 
curves did not decline as the exponential disk of starlight would predict. 
This was quickly followed by observed flat rotation curves beyond the
optical disks by workers as
Mort Roberts, Albert Bosma and others at 21 cm, and by Vera Rubin and others
in radial velocities of HII-regions within the optical boundaries.
The presence of
a halo of dark matter that was assumed to be more or less
spherical. Not much later it was realised that for galaxies to remain stable, 
an additional dark matter component was required. Analysis of the thickness of 
the \hi layer in NGC\,891 \citep{vdk81c}, showed that this dark matter could 
not reside in the disc and would form a more or less spherical structure
in which the visible galaxy would be embedded. These developments
have been discussed by many authors, e.g. \citet{OBrien2010A}, 
\citet{kf11} and \citet{JBH13}.

To this date however, the nature of the dark matter remains elusive.
By understanding the dark matter halo in detail, it will be possible for 
physicists to place stronger constraints on the theoretical nature of the dark 
matter. In this series of papers, which is a major part of the PhD 
thesis of the first author, \citet{peters14}, we  continue the work by 
\citet{OBrien2010A} and \citet{OBrien2010B,OBrien2010C,OBrien2010D}, who 
set out to measure the shape of the dark matter halo around edge-on galaxies
using the flaring of the \hi layer.

Because of their side-on orientation, edge-on galaxies offer a unique 
perspective on the structure of galaxies.
Due to the random orientation of galaxies in space, only a fraction is 
seen edge-on. 
Our Galaxy can be considered a prime example of an edge-on galaxy.
There are many advantages to the edge-on perspective.
For example, the longer lines-of-sight allow the study of the baryonic 
content at very low volume densities.
This led \citet{vdk79} to the discovery that most stellar discs have a 
truncation in their outer regions, beyond which the stellar density rapidly 
drops to zero. In face-on galaxies only modern techniques have been able
to establish the existence of truncation, see \citet{Peters2015G} 
and \citet{peters14}.

Another advantage of the edge-on perspective is that they allow for the study 
of the structure of the discs as function of height above the plane.
This advantage led to the discovery of thick stellar discs, using surface 
photometry. Evidence for this additional mass component has been 
detected first in S0 galaxies \citep{Burstein1979A,Tsikoudi1980A}.
in the Galaxy, and then
\citet{Gilmore1983A} using stellar counts.
In face-on galaxies, the vertical direction is integrated along the 
line-of-sight, making the thick and thin disc hard to disentangle.

While offering a unique perspective on the vertical structure of the disc, 
each line-of-sight also forms a superposition of light emitted at 
various radii, as the line crosses through the disc. 
Decomposition of the stellar discs is required before the galaxy can be 
understood as function of radius $R$ and height $z$.
Many authors have thus developed new and ever more sophisticated methods to 
disentangle this information 
\citep[e.g.][]{vanderKruit1981A,Pohlen2007a,Comeron2011A}.
The most modern methods are able to model the full spectrum from near 
infrared to optical and can account for the dust reprocessing of the 
light \citep{Geyter2013A}.

For neutral hydrogen, the situations is even more complex. 
Where most optical and infrared emission comes in the form of very broad 
lines and blackbody emission, the hydrogen line is very narrow.
Because of this, the most prominent source of broadening of the \hi line 
emission of a galaxy is due to the Doppler shift from the bulk motion of the 
gas, in particular due to the rotation around the galaxy and the turbulence of 
the gas.
The broadening due to the turbulence of the gas (also known as the velocity 
dispersion) is typically of order 10\,km/s and is low compared to the 
rotational velocity of the galaxy, which is typically of order 100\,km/s
or more.
If the observation of an edge-on galaxy is well resolved spatially and in 
velocity, for example using a radio interferometer, the line-of-sight 
velocity $v_\textrm{los}$ offers an additional constraint on a decomposition.

We begin in this first paper of the series with an analysis of the neutral 
hydrogen (\hi) content of eight edge-on galaxies.
One of the main questions that arises in this paper is the possible impact 
that \hi self-absorption of the gas can have on any subsequently derived 
property.
Traditionally this topic has largely been ignored, but we argue that it has to be 
addressed before any sensible statement can be made about the dark matter halo
 from \hi data.
The currently available modelling tools are unable to treat for \hi 
self-absorption, which is why we introduce a new software tool called 
\textsc{Galactus} in paper II. 
This tool allows the users to perform a (Bayesian) fit to the 
HI data cubes.
We use this tool to model the face-on galaxies NGC\,2403. 
Projecting the results to an edge-on model, we find that the maximum 
temperature maps from this paper should indeed have been 
higher, concluding that self-absorption indeed is important.
We also demonstrate how the missing mass fraction due to the \hi 
self-absorption increases with inclination.

In paper III, we return to the radio cubes presented here. 
Using a series of simulated \hi cubes, we demonstrate that with our new 
software it is possible to analyse the edge-on galaxies. 
We also demonstrate how treating the \hi as optically thin can lead to 
errors in the measured properties.
The rest of the paper is dedicated to analysing the eight edge-on galaxies.

Paper IV is dedicated to modelling the stellar and dust contents of 
these edge-on galaxies.
We use the software package \textsc{FitSKIRT} to model the galaxies in 
multiple wavelengths using optical and near-infrared observations.

In paper V, we use these \hi and stellar results to model the theoretical 
hydrostatics of the edge-on galaxies.
Comparing the theoretical and observed hydrostatics enables us to measure the 
gravitational potential of the dark matter halo in both the radial and 
vertical direction.
We fit dark matter haloes to the best five out of our initial eight galaxies.
Three galaxies have dark matter haloes that are nearly spherical, while two 
more have strongly prolate haloes.
We did not find any strongly oblate haloes.

\section{Introduction to this paper}

The first spatially resolved \hi observations of edge-on galaxies 
came in the late seventies, but lacked sufficient quality 
for a detailed analysis \citep{Sancisi1976A, Weliachew1978A}.
That became possible with the subsequent observation of NGC\,891 by 
\citet{Sancisi1979A}, a nearby edge-on galaxy similar to the Galaxy 
\citep{vanderKruit1984A}.
Sancisi \& Allen modelled both the rotation curve and the thickness 
of the disc, concluding that the thickness was less than one kpc in 
the inner parts, but flaring out to 1-2\,kpc thickness at 20-24\,kpc radius. 
It was shown by \citet{vdk81c} that the thickness of this 
\hi layer should and does increase exponentially with radius, which was later 
confirmed by \citet{Rupen1991A}.
Since then, NGC\,891 has remained a prime target for subsequent 
observations, each time improving the quality of the data and revealing new 
structure, such as the presence of an \hi well below and above the disc
\citep{Swaters1997A,Oosterloo2007A}. {We also note that this gas 
in NGC\,891 is lagging behind the \hi in the disk. We refer also to the work
of \citet{Matthews2003A} on one of the galaxies in our sample, UGC\,7321. 
These important developments produce effects that will have to be kept in mind.

\citet{Olling1996A} was the first to derive the radial \hi surface density 
of an edge-on, by decomposing the galaxy using inverse Abel transformation 
in a technique inspired by the work on face-on galaxies by \citet{Warmels1988A}.
While the velocity dispersion in face-on galaxies can be measured more or 
less directly, the superposition of gas 
in edge-on systems at various radii requires a careful 
decomposition of the galaxy.
In the same paper, \citet{Olling1996A} also used the outer envelope of the 
XV-diagram to measure the velocity dispersion variation in NGC\,4244, which 
was found to be consistent with a velocity dispersion of roughly 8.5\,km/s.
\citet{OBrien2010C} modelled the full XV-diagram for eight edge-on galaxies 
and found that most systems display \hi velocity dispersions of 6.5 to 
7.5\,km/s and all except one show radial structure in this property.

We ultimately 
aim to measure the shape of the dark matter halo around edge-on galaxies.
By carefully modelling the theoretical hydrostatics of the neutral hydrogen 
and stellar components in each galaxy, and comparing this with the observed 
hydrostatics, it is possible to discern the effects due to the dark matter halo.
This is a continuation of the project started in 
\citet{OBrien2010A,OBrien2010B,OBrien2010C,OBrien2010D}, who modelled these 
components for galaxy UGC\,7321 and found a near spherical halo.
In this paper, we present the \hi observations for eight edge-on galaxies, 
similar to the sample of \citet{OBrien2010A}. 
A key question from \citet{OBrien2010A} concerns the possible effects of 
self-absorption of the neutral hydrogen.
We begin this paper by re-addressing this issue.

\begin{figure*}
\centering
\includegraphics[width=65mm]{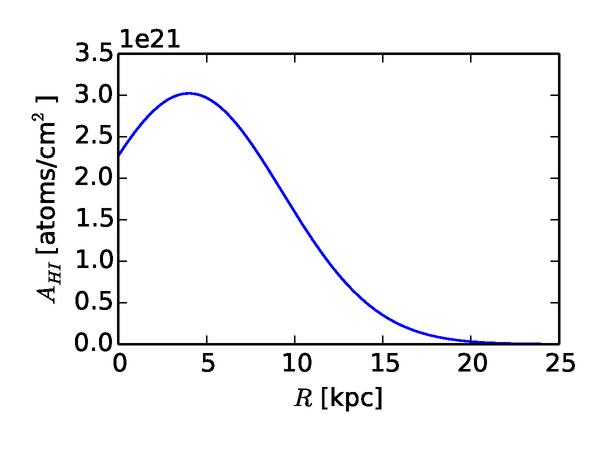}
\includegraphics[width=65mm]{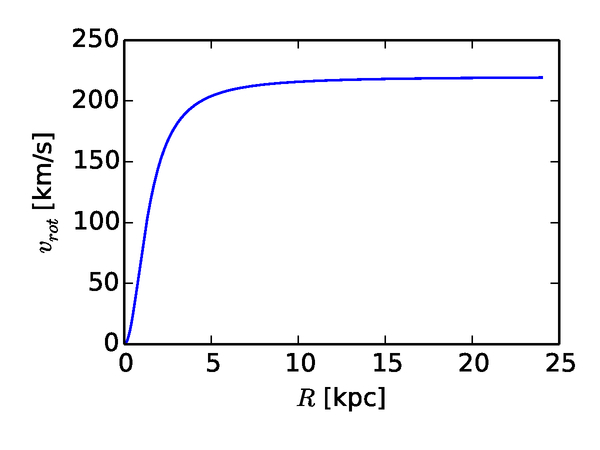}
\includegraphics[width=62mm]{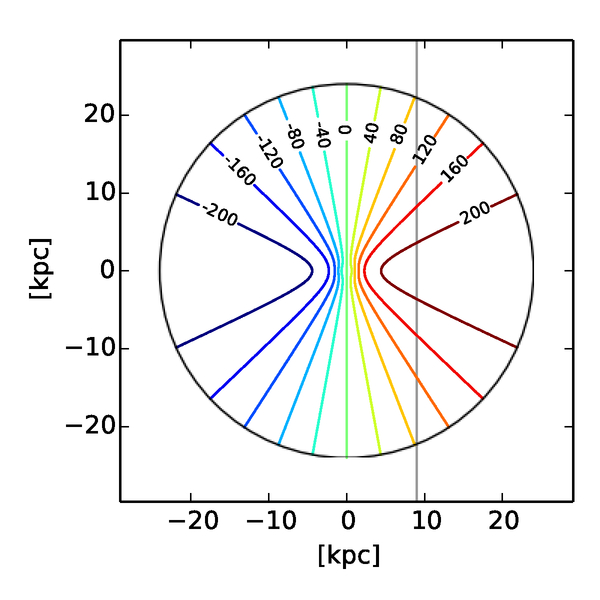}
\includegraphics[width=65mm]{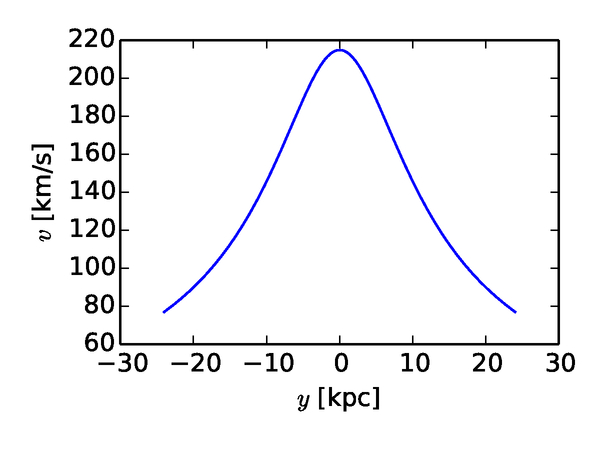}
\includegraphics[width=65mm]{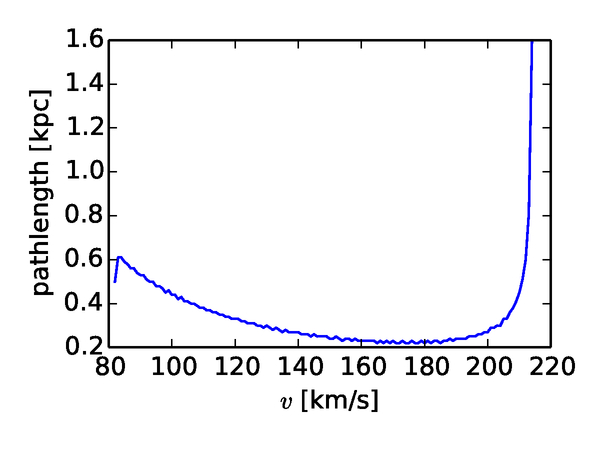}
\includegraphics[width=65mm]{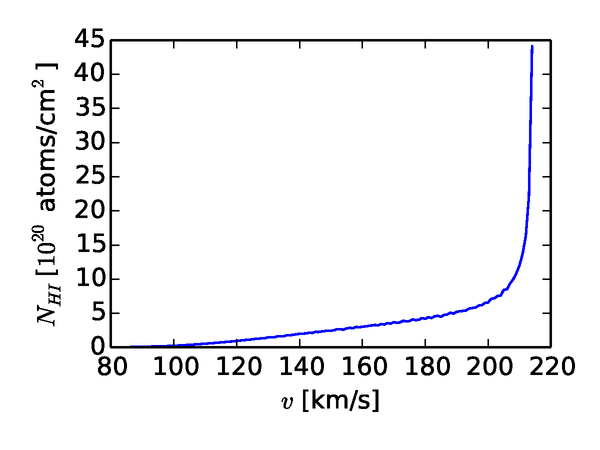}
\includegraphics[width=65mm]{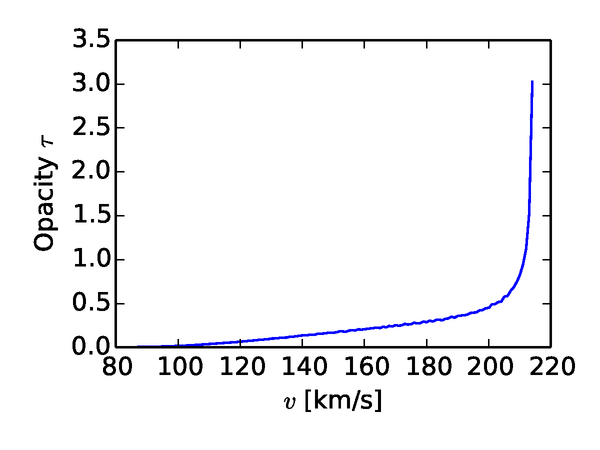}
\includegraphics[width=65mm]{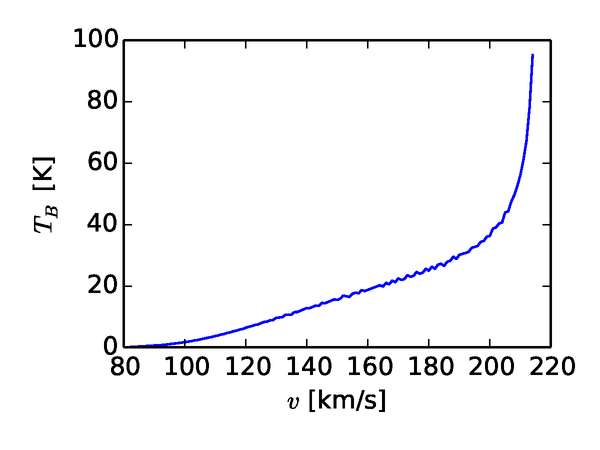}
\caption[Tracing a ray through a toy model]{Tracing a ray through a toy model. Top row: Face-on surface density and rotation curve of the toy model. Second row left: Spider diagram of the projected velocity at each position in the disc, with the grey vertical line demonstrating the ray at nine kpc. Second row right: projected velocity along the ray. Third row left: The distance crossed by the ray in each projected velocity. Third row right: The amount of \hi atoms at each projected velocity. Bottom row left: Absorption coefficient $\tau_\nu$ with velocity. Bottom row right: Observed brightness with velocity.}\label{fig:toymodel}
\end{figure*}

\begin{table*}
\begin{minipage}{126mm}
\begin{tabular}{lll}
Parameter & Function & Value\\\hline\hline
Inclination & & $i=90^\circ$\\
Thickness of disc & & FWHM$=700$\,kpc\\
Systemic velocity & & $v_\textrm{sys}=0$\,km/s\\
Rotation curve & $v_\textrm{rot}(R) = v_\infty(1 - \frac{1}{R^2/a^2 + 1})$ & $v_\infty=220$\,km/s \\
& & $a=1.4$\,kpc\\
Face-on surface density & $A_\hi(R) = A_\textrm{max}  \exp\left(\frac{(R-\mu)^2}{2\sigma^2}\right)$ & $\rho_0=1.4$\,atoms/cm$^3$\\
& & $A_\textrm{max} = 3\times10^{21}$\,atoms/cm$^2$\\
& & $\mu=4.0$\,kpc\\
& & $\sigma=5.3$\,kpc\\
\hline\hline
\end{tabular}
\end{minipage}
\caption[Parameters of the toy model]{Parameters used in the toy model}\label{tbl:toymodel}
\end{table*}

\begin{figure*}
\centering
\includegraphics[width=84mm]{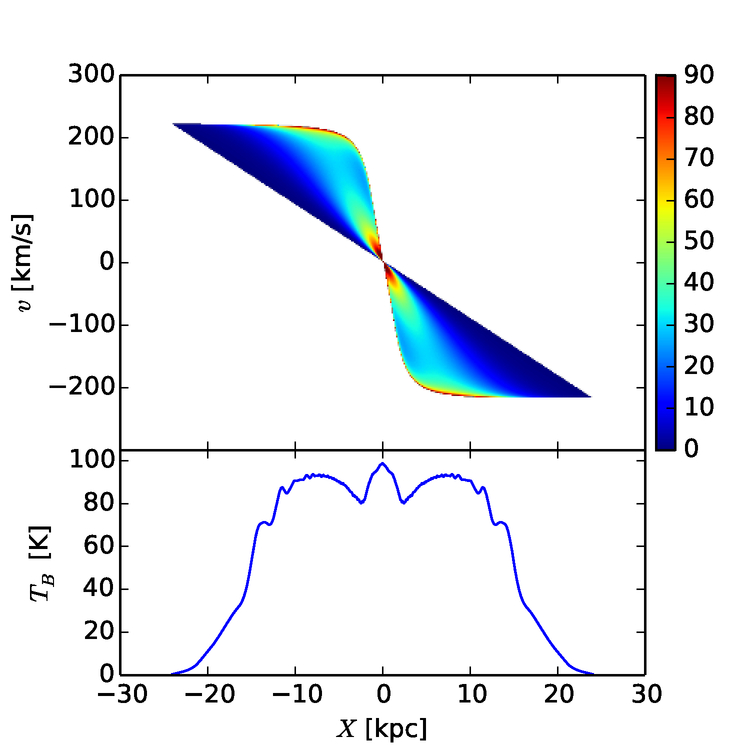}
\includegraphics[width=84mm]{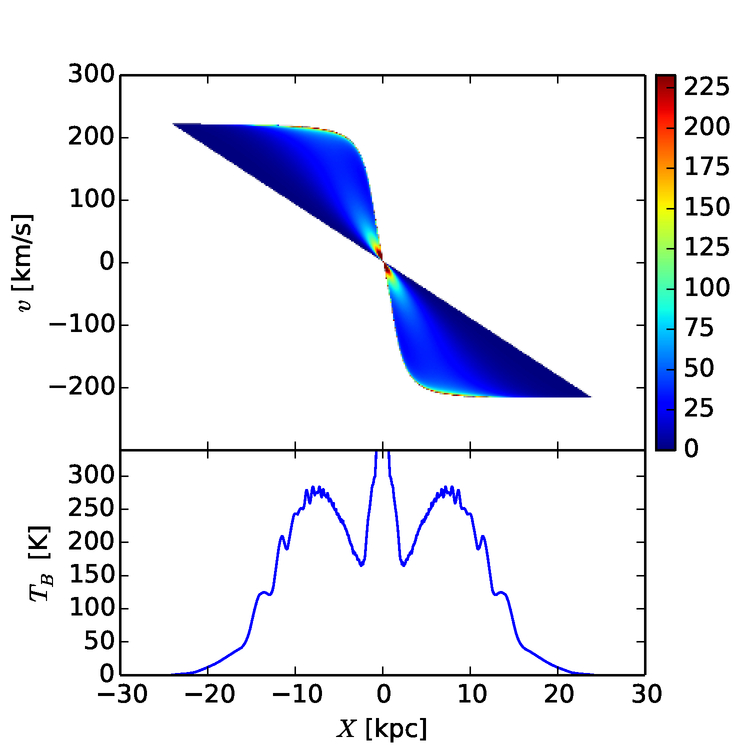}
\caption[XV-diagrams for the toy model]{Position-velocity diagrams for the toy model. The vertical bars scow the scale of brightness temperature.} The left panel shows results for a self-absorption model. The right panel shows the result for an optically thin model. The lower plot in both panels shows the maximum brightness temperature at each position in the XV-diagram\label{fig:toymodel-PV}
\end{figure*}

\section{A Toy Model}\label{sec:toymodel}
A continuing concern throughout this series of papers 
will be the role of the \hi 
self-absorption (sometimes referred to as HISA in literature) on the 
observations.
In our own Galaxy, many lines of sight are known to become optically thick 
within several hundred parsec, with faint HISA features visible in all 
direction where the background emission is sufficiently strong, plus stronger 
HISA features forming distinct cloud complexes 
\citep{Gibson2005A,Draine2011,Allen2012A}.
Such cloud complexes cannot be resolved in other galaxies.
Most papers on external galaxies therefore treat the \hi as if it is optically 
thin.
Evidence is however accumulating that flat-topped \hi profiles characteristic 
of this effect are common in M31 and M33, where hydrogen mass correction 
factors of order 1.3 to 1.4 have been derived \citep{Braun2012A}. 
So what is the overall opacity of the \hi in external galaxies and should we 
correct for it?
One answer comes from \citet{OBrien2010A}, who show maps of the maximum 
surface brightness for eight edge-on galaxies.  
Rather than in the conventional unit of mJy per beam, the data are presented 
in Kelvin. 
A remarkable result is found: In each galaxy, the maximum brightness along 
the major axis is roughly constant between 80 and 100\,K along most of the 
disc.
This is most clear in ESO\,274-G001, of which a similar result is shown here 
in Figure \ref{fig:ESO274-G001} top-right.
While the path-length through the disc varies drastically with position along 
the major axis, why can the maximum surface brightness remain so constant?
\citet{OBrien2010A} speculated that this might be due to \hi self-absorption.
If true, this is a very important result: \hi should not be treated as 
optically thin.

\begin{table*}[t]
\centering
\resizebox{0.98\textwidth}{!}{
\begin{tabular}{llllllcccc}
\textbf{Galaxy}      & \textbf{RA} & \textbf{DEC} & \textbf{PA} & \textbf{Dist} && \textbf{Scale} & \textbf{Morphology} & \textbf{$M_B$}\\
\textbf{ }           & \textbf{ } & \textbf{ } & \textbf{[$^\circ$]} & \textbf{[Mpc]} && \textbf{[pc/arcsec]} & & \textbf{[B-mag]}\\
                     &     (1)     &     (1)      &  (1)  & (2)   & (2)     &      & (3) & (4)   \\*\hline\hline
{IC\,2531}      &   9:59:56.0 & -29:37:00.0 &  74.6 & 27.2 &TF (a)   & 131.7& Sc    &  8.835 (b)  \\*
{IC\,5052}      &  20:52:05.5 & -69:12:04.4  & 321.2 &  5.7 &TRGB (b) & 27.4 & SBd   &  9.102 (b)  \\*
{IC\,5249}      &  22:47:06.0 & -64:50:00.9  &  14.4 & 32.1 &TF (a)   & 155.4& SBd   & 12.314 (a)  \\*
{ESO\,115-G021} &   2:37:46.9 & -61:20:13.2  & 223.2 &  4.9 &TRGB (a) & 23.8 & SBdm  & 14.301 (a)  \\*
{ESO\,138-G014} &  17:07:02.2 & -62:05:20.9  & 134.8 & 15.8 &TF (a)   & 76.5 & SB(s)d& 10.663 (b)  \\*
{ESO\,146-G014} &  22:12:59.9 & -62:04:06.3  & 222.3 & 21.7 &TF (a)   & 105.1& SBd   &   -         \\*
{ESO\,274-G001} &  15:14:14.3 & -46:48:21.0  & 216.6 &  3.0 &TRGB (a) & 14.6 & SAd   & 8.416 (b)   \\*
{UGC\,7321}     &  12:17:34.2 & +22:32:25.6  &   8.1 & 10.0 &TRGB (c) & 48.5 & Sd    & 10.883 (b)  \small\\*\hline\hline
\end{tabular}}
\caption[Properties of the edge-on sample]{General properties of the galaxies in this survey.
   Distances are based on the Tully-Fisher relationship (TF) or the tip of the Red Giant 
Branch (TRGB).
  References: (1): This work, (2a): \citep{Tully2008A}, (2b): \citep{Radburn-Smith2011A}, 
(2c): \citep{Matthews2000A}, (3): \citep{Vaucouleurs1991A}, (4a): \citep{Skrutskie2003A} or 
(4b):\citep{Jarrett2003A} }
\label{tbl:opticalproperties}
\end{table*}

We begin by analysing a toy model of an edge-on galaxy. 
We roughly base our model on the observations of NGC\,891 by 
\citet{Sancisi1979A}.
NGC\,891 bears close resemblance to our Galaxy \citep{vanderKruit1984A}.
Because of this, we use the Solar Neighbourhood average \hi density 
$\rho_0$ of 1.4 atoms/cm$^3$ \citep[Table 9-1]{Mihalas1981}, as this 
provides a reliable estimate of the density.
Rather than the value of 1\,kpc adopted by \citet{Sancisi1979A} for the
constant thickness FWHM of the disc, we use 700\,pc to compensate for the 
revised distance of NGC\,891 to 10\,Mpc.
The \hi forms a smooth medium of constant density with height and only changes 
with radius, as defined through the face-on surface-density distribution.
We use the average density $\rho_0$ and thickness FWHM to set the maximum 
face-on surface density $A_\textrm{max}$ at $3.0\times10^{21}$\,atoms/cm$^3$.
The face-on surface density and rotation curve of the toy model are shown in 
the top row of Figure \ref{fig:toymodel}, with the exact parameters shown in 
Table \ref{tbl:toymodel}.
The disc has a maximum radius of 24\,kpc.
The gas in our model has no intrinsic dispersion, so all emission occurs at 
exactly the projected velocity $v_\textrm{los}=v_\textrm{rot} \cos(\theta)$. 
Here $\theta$ denotes the angle between the position of the gas ($x,y$) inside 
the galaxy, and the centre of the galaxy [$\theta=\arctan(y/x)$].
The lack of intrinsic dispersion is similar to a velocity dispersion lower 
than one km/s.
This is not realistic and would cause an overestimation of the self-absorption.
To correct for this, we divide our calculated
\hi absorptions by a factor 8, which is 
\emph{very} roughly the same effect as expected due to a velocity dispersion 
of 8 km/s. 
\citet{Lewis1984} studied 200 face-on galaxies and found the velocity 
dispersion to be $8\pm1$\,km/s, which is consistent with observations 
of the outer Galaxy \citep{Saha2009}.
The results presented should thus be seen as a lower limit to the 
self-absorption.
We do not include any form of beam smearing.

We start by tracing a single ray through the disc at a position 
$X$ along the major axis of nine kpc from the centre.
The position of the ray is shown in the spider diagram in Figure 
\ref{fig:toymodel}.
As the ray passes through the model, it crosses through various regions, each 
of which is at a particular projected velocity $v_\textrm{los}$. 
To illustrate this, we also show the projected velocity along the ray in 
Figure \ref{fig:toymodel}, at the right, second panel from the top.
As is clear from that panel, the velocities are not evenly distributed along 
the ray.
This means that the distance that the ray crosses --also known as the 
path-length-- is different for each projected velocity $v_\textrm{los}$.
This is demonstrated in the fifth panel of Figure \ref{fig:toymodel}, where we 
show the path-length per velocity.
Near the lowest and highest velocities, the ray crosses the longest 
path-lengths. 
Not only has the velocity changed along the ray, but also the amount of atoms. 
While the path-lengths at the low velocities are the longest, there is also 
the least amount of \hi there.
The highest observed velocity (the terminal velocity) is near the inner 
radii of the model, and as such harbors the most neutral hydrogen.
This is demonstrated in the sixth panel of Figure \ref{fig:toymodel}, 
where we show the amount of  $N_{H_I}$ (in atoms/cm$^2$) per projected 
velocity $v$.
Using Equation \ref{eqn:tau2}, we can subsequently convert the amount 
of \hi to the absorption coefficient $\tau_\nu $ \citep[combination of 
Equations 8.8 and 8.11 from][]{Draine2011},
\begin{equation}
 \tau_\nu =  2.190\, \sqrt{2\pi} \,\frac{N_{H_I}}{10^{21}} \frac{100\textrm{K}}{T_\textrm{spin}}\,\,. \label{eqn:tau2}
\end{equation}
The spin temperature $T_\textrm{spin}$ is chosen at 100 Kelvin.
The result is shown in Figure \ref{fig:toymodel}, where it can be seen 
that the absorption coefficient is roughly 0.4 throughout most of the 
velocity span, yet spikes to over three near the terminal velocity. 
The terminal velocities and thus the inner radii of the model have 
turned opaque.
This result is similar to the (much more detailed) simulation of the 
Galaxy by \citet{Douglas2010}, who showed that a large fraction of the 
Galactic plane should be optically thick.
Using Equation \ref{eqn:Tgood} we can convert the opacity to an observed 
brightness \citep[][Equation 7.26]{Draine2011}, which we show in the last 
panel of Figure \ref{fig:toymodel},
\begin{equation}
 T_B = T_\textrm{spin}\left(1 - e^{-\tau_\nu}\right)\label{eqn:Tgood}\,\,.
\end{equation}
As can be seen, the observed brightness along this ray rises continuously 
towards the maximum of 100\,K at the terminal velocity.

\begin{table*}
\centering
\resizebox{0.98\textwidth}{!}{
\begin{tabular}{l|ccccc|ccccc}
 & \multicolumn{5}{c}{\textbf{High Resolution}} & \multicolumn{5}{c}{\textbf{Low Resolution}}\\
\textbf{Galaxy} & \textbf{$\theta_\textrm{FWHM}$} & \textbf{$\Delta $v} & \textbf{$\sigma$} & \textbf{$\sigma$} & \textbf{$\sigma$} & \textbf{$\theta_\textrm{FWHM}$} & \textbf{$\Delta $v} & \textbf{$\sigma$} & \textbf{$\sigma$} & \textbf{$\sigma$}\\*
& \textbf{[arcsec]} & \textbf{[km/s]} & \textbf{[$\frac{\textrm{mJy}}{\textrm{beam}}$]}  & \textbf{[K]} & \textbf{[atoms/cm$^2$]}& \textbf{[arcsec]} & \textbf{[km/s]} & \textbf{[$\frac{\textrm{mJy}}{\textrm{beam}}$]} & \textbf{[K]} & \textbf{[atoms/cm$^2$]}\\*
\hline\hline
IC\,2531 & 10.0 & 6.6 & 1.2 & 7.1 & $8.5 \times 10^{19}$ & 30.0 & 6.6 & 1.5 & 1.0 & $1.2 \times 10^{19}$ \\
IC\,5052 & 11.0 & 3.3 & 1.4 & 6.8 & $4.1 \times 10^{19}$ & 30.0 & 3.3 & 1.7 & 1.1 & $6.6 \times 10^{18}$ \\
IC\,5249 & 8.0 & 6.6 & 1.1 & 10.8 & $1.3 \times 10^{20}$ & 30.0 & 6.6 & 1.9 & 1.3 & $1.5 \times 10^{19}$ \\
ESO\,115-G021 & 10.6 & 3.3 & 1.2 & 6.6 & $3.9 \times 10^{19}$ & 30.0 & 3.3 & 1.4 & 0.9 & $5.5 \times 10^{18}$ \\
ESO\,138-G014 & 11.8 & 6.6 & 1.7 & 7.3 & $8.7 \times 10^{19}$ & 30.0 & 6.6 & 2.3 & 1.6 & $1.9 \times 10^{19}$ \\
ESO\,146-G014 & 8.6 & 6.6 & 1.2 & 9.7 & $1.2 \times 10^{20}$ & 30.0 & 6.6 & 1.7 & 1.1 & $1.4 \times 10^{19}$ \\
ESO\,274-G001 & 12.2 & 3.3 & 1.5 & 6.2 & $3.7 \times 10^{19}$ & 30.0 & 3.3 & 1.8 & 1.2 & $7.2 \times 10^{18}$ \\
UGC\,7321 & 13.8 & 5.2 & 0.3 & 1.0 & $9.8 \times 10^{18}$ & 30.0 & 5.2 & 0.5 & 0.4 & $3.4 \times 10^{18}$ \\
\hline\hline
\end{tabular}}
\caption[Global properties of the \hi cubes]{\label{tbl:GlobalPropertiesHICubes}
The global properties as measured from both the high and low-resolution \hi 
cubes. Shown are the size of the beam $\theta_\textrm{FWHM}$, the width and 
stepping of the velocity channels $\Delta v$ and the one-sigma background 
noise expressed in various units.}
\end{table*}

We trace rays at each position along the major axis of the disc. 
We then build a position-velocity (XV) diagram, such as can be seen in the 
top part of Figure \ref{fig:toymodel-PV}. 
In the right panel of Figure \ref{fig:toymodel-PV} we demonstrate the 
equivalent result, as calculated using the optically thin conversion 
($\tau_\nu \simlt 0.1$) from $N_\hi$ to $T_B$ given by Equation 
\ref{eqn:Tbad} \citep{OBrien2010A},
\begin{equation}
 T_B = \frac{N_{H_I}}{1.82\times10^{18}\, dv}\,\,.\label{eqn:Tbad}
\end{equation}
The scales in both cases encompass the full range in intensity.
Comparing both panels, we see that near the centre the diagrams are very 
similar, but over the rest of the diagram there is clearly a large difference 
between the self-absorbing and optically thin models. 
In the self-absorption model, a bright bar forms near the terminal velocities. 
This is due to the stacking of gas at more outward positions and lower 
projected velocities.
In the optically thin model, this gas is always visible, and any envelope 
tracing method will have no problem understanding it.
However, the stacking is a problem in the self-absorption model, as any 
outer envelope tracing method will interpret the diagram as less \hi at 
the more outward radii than is actually there.

This is further illustrated by the lower panel of both figures, where we 
show the maximum brightness that the ray obtains at that position.
The self-absorbing model shows a plateau between 80 and 100\,K, much the 
same as the results by \citet{OBrien2010A}. 
In contrast, the optically thin model does not show such a plateau at 
all\footnote{Note that the bumps in the profile are due to the finite 
numerical resolution}. 
The observed brightness temperature rises well above 200\,K in the inner part.
Larger regions with brightness levels well above 100\,K are never observed 
in reality. 
The maximum brightness peak in \citet{OBrien2010A} is still below 200\,K.  
Even in the high-resolution observations available by studying our own Galaxy, 
temperatures above 150\,K are uncommon \citep{Taylor2003,Douglas2010}.

So what is the consequence of this?
As demonstrated by this toy model, the presence of a maximum brightness 
plateau along most of the major axis, as well as the limited brightness, 
are both strong indications of \hi self-absorption. 
We conclude that any subsequent modelling will need to take into account 
this effect, or risk underestimating the \hi content of the galaxies.

\section{Sample \& Observations}\label{sec:SampleNObservations}

Our sample is the same as that of \citet{OBrien2010A}.
Criteria for the galaxies were that they lie in the southern sky, are close to 
edge-on
($a/b \geq 10$), Galactic latitude $\|b\| \geq 10^\circ$ as to avoid
optical and infrared extinction by the Galaxy, relatively bulge-less (Hubble 
type Scd-Sd),
nearby enough to be able to resolve the flare ($d \leq 30$ Mpc) and with a 
minimal
integrated flux of 15 Jy\,km/s.
This led to a sample of five edge-on galaxies.
Two other galaxies were added for various reasons.
{UGC\,7321} was added due to the availability of very high quality archival 
data,
from the work by \citet{Uson2003A}.
{ESO\,274-G001}, despite its low Galactic latitude of $9.3^\circ$, was
included for its exceptional proximity.
{IC\,2531}, a box/peanut bulge, barred galaxy was included as a test of
halo shapes under different types of mass scales and stages of secular
evolution.
We list the global properties of the sample in Table
\ref{tbl:opticalproperties}.
We refer the reader to Paper IV for more optical and 
near-infrared images of these galaxies.

The results presented here come from both archived data as well as new 
observations.
Almost all were observed using the Australian Telescope Compact Array (ATCA) 
near Narrabri, Australia.
The only exception is UGC\,7321, which was based on archived observations 
from the
Very Large Array (VLA). \citet{OBrien2010A} also included ATCA observations 
for this galaxy, but we found that this limited 4 hour observation reduced 
the overall quality of the cube.
Most of the observations came from previous archived observations by various 
authors over a range of years, going back to 1993.
A significant number of the observations was already used by 
\citet{OBrien2010A}.
A major new set of observations came as part of the work undertaken by The 
Local Volume \hi Survey (LVHIS) group \citep{Koribalski2008}.
For {ESO\,138-G014} and {ESO\,274-G001}, we have taken new observations 
using the ATCA.
The various observations and configurations are listed in Table 
\ref{tbl:HIobservations}.
In total, we have collected more than 132 hours of additional 
observation-time compared to \citet{OBrien2010A}.

\section{Reductions}\label{sec:HIreductions}

\subsection{Calibration \& Imaging}\label{section:HIreduction}
The calibration and imaging of the data has been done in \textsc{miriad}
\citep{Sault1995A}. All subsequent analysis was done in \textsc{python}.
All archived data from the ATCA was take in the XX and YY polarizations.
The new observations also include the XY and YX polarizations.
We made use of the pre-calibrated data from \citet{OBrien2010A}, although we
have re-reduced parts where we believed improvements could be made.

The data were first imported and split into the various observations, excluding
any telescope shadowing. We then inspected all data for RFI and flagged where
necessary. The primary and secondary calibrator were then used to calculate the gains and
band-passes. This was then applied to the main target. The continuum was then
subtracted from the UV data.

The imaging was done using a three-pass strategy, similar to
\citet{OBrien2010A}. 
Since the exact pointing sometimes varied between observations, we used the 
\textsc{miriad} ``joint approach'' to invert the data in mosaic
mode. 
The data were imaged in Stokes II, which is a special case of Stokes I,
under the assumption of an unpolarised source\footnote{From the 
\textsc{Miriad} user manual at {http://www.atnf.csiro.au/\linebreak[4]
computing/software/miriad/userguide/node108.html}}. 
A robust parameter of 0.4 was used to
provide a good balance between beam-size and side-lobes.

In the first pass, we cleaned the full cube, using the maximum-entropy
\textsc{mosmem} task\footnote{There is an ongoing debate on the merits of 
using clean versus maximum entropy. We have chosen the maximum entropy method, 
as the \textsc{miriad} user guide recommends this as the superior method to 
use, in particular for extended emission.}. This was then restored, which 
resulted in an estimate for the
major and minor axis of the beam. This was taken as the basis for the second
pass. Here we used a square cell-size of half the full-width half-maximum 
(FWHM) of the major axis beam. The central quadrant was again
cleaned and restored, using a circular beam with the FWHM of the major axis 
of the beam from the
previous pass. The restored cube was smoothed to twice the FWHM and the 
standard deviation $\sigma$ of the background
noise  determined. We defined a regions mask in the cube where the signal 
strength was more than double the noise. In the third pass, the data raw cube 
was again cleaned with \textsc{mosmem}, this time only using the new regions 
mask.

We also produced low-resolution cubes with a beam of $30''$ to search for any 
faint,
extended component. This was done by tapering the Fourier transformation to 
$30''$. The clean
was then based on the same regions as the high-resolution cube. In some cases we
found the regions too small for the low-resolution beam, such that parts of 
the galaxy were cut off. The regions were
then widened and both the high- and low-resolution cubes were again cleaned 
and restored using these new regions.

The VLA data for {UGC\,7321} consisted of very small band-passes, such that 
there
was no possibility to perform a continuum subtraction in the UV-plane. The 
continuum
was therefore subtracted in the image plane. The above strategy led to a 
beam of 
15 arcsec for UGC\,7321. We have applied an additional side-lobe suppression 
of 700 arcsec
to the beam, which led to the smaller beam as reported here.

For IC\,2531, IC\,5249, ESO\,138-G014 and ESO\,146-G014, the channel-width and 
spacing was originally 3.298\,km/s. 
However, since these galaxies are the furthest, they also have the highest 
pc/arcsec scales (see Table \ref{tbl:GlobalPropertiesHICubes}), which meant 
that the original beams were not fully resolving the inner disc.
We have therefore doubled the channel-width and spacing, lowering the noise 
and thus enabling us to decrease the size of the beam, such that all have a 
FWHM beam of $\sim1$\,kpc.
This failed for IC\,2531, where the effective noise was too high for any 
practical purpose.
The beam for that galaxy has now been set at 1.3\,kpc.
A similar problem occurred for IC\,5249, where we now have set the beam to a 
FWHM of 1.25\,kpc.

\subsection{Analysis}

\begin{table*}
\centering
\begin{tabular}{l|cccccc}
\textbf{Galaxy} & \textbf{v$_\textrm{sys}$} & \textbf{v$_\textrm{max}$} & \textbf{$W_{20}$} & \textbf{$W_{50}$} & \textbf{FI} & \textbf{$M_\textrm{\hi}$}\\*
  & \textbf{[km/s]} & \textbf{[km/s]} & \textbf{[km/s]} & \textbf{[km/s]} & \textbf{[Jy km/s]} & \textbf{[M$_\odot$]}\\*\hline\hline
{IC\,2531}      & 2463.8 & 260.5  & 484.8 & 468.3 & 41.7 & $7.3 \times 10^{9}$\\*
{IC\,5052}      & 596.7  & 114.4  & 197.9 & 181.4 & 117.7 & $8.9 \times 10^{8}$\\*
{IC\,5249}      & 2351.7 & 131.9  & 234.2 & 217.7 & 23.2 & $5.6 \times 10^{9}$ \\*
{ESO\,115-G021} & 514.2  &  85.1  & 141.8 & 125.3 & 109.0 & $6.2 \times 10^{8}$ \\*
{ESO\,138-G014} & 1492.3 & 130.9  & 237.5 & 227.6 & 48.9 & $2.9 \times 10^{9}$ \\*
{ESO\,146-G014} & 1678.8 &  84.1  & 148.4 & 128.6 & 17.5 & $1.9 \times 10^{9}$ \\*
{ESO\,274-G001} & 523.6  & 103.9  & 181.4 & 168.2 & 150.1 & $3.2 \times 10^{8}$ \\*
{UGC\,7321}     & 409.3  & 128.8  & 237.0 & 221.6 & 41.7 & $9.8 \times 10^{8}$ \small\\*\hline\hline
\end{tabular}
\caption[Spectral properties of the edge-on galaxies]{Spectral properties of the galaxies}
\label{tbl:HISpectrumProperties}
\end{table*}

The noise in a mosaicked cube varies with position. It is therefore important
to measure the noise as close to the galaxy as possible, while avoiding any 
contamination from
the galaxy. The region-mask offered a good tool for this. 
We combined all region-masks into a single master-mask, which defined the 
maximum extent of the region-masks. The noise was then measured on the pixels 
in every channel that
were part of the master mask, but were outside the specific region of
that channel. We show the noise estimates in Table 
\ref{tbl:GlobalPropertiesHICubes}. 
Following the results from Section \ref{sec:toymodel}, we present 
all results in Kelvin, as this is the most natural unit in which 
we can identify self-absorption.
We convert from flux density $S$ in Jy/beam to surface brightness 
temperature $T_B$ in Kelvin using Equation \ref{eqn:JytoK}, and to 
column densities $N_\textrm{\hi}$ in atoms/cm$^2$ using the optically thin 
Equation \ref{eqn:Tbad},
\begin{equation}
 T_B = \frac{\lambda^2 S}{2 k_B \Omega} = \frac{606000 S}{\theta^2}\,\,.\label{eqn:JytoK}
\end{equation}

The central position and position angle (PA) were derived from the 
high-resolution cubes. First, the cube was clipped below 30\,K, as
the lower intensity regions were often found to be asymmetrical. A 
clear example
of this asymmetry can be seen in IC\,5052 (see Figure 
\ref{fig:IC5052} bottom-left).
We then created a moment-zero image of the cube.
The centre of the galaxy was found as the minimal $\chi^2$ of the
difference between the image and its $180^\circ$ rotated counterpart, as trialed
over a wide range of positions. The position angle was then found by rotating
the galaxy over a range of possible angles around this central position. The
lowest $\chi^2$ between the rotated images and its upside-down flipped
counterpart gave the position angle. 
We define the position angle such that the receding side always becomes the 
left side, when rotating the image.
The results for this are accurate to within $0.1$ pixel and $0.1$ degree and 
are listed in Table \ref{tbl:opticalproperties}.
These positions will only serve as first-guess positions for the kinematic 
fitting in Paper III, and are thus not 
particularly important.

Both the high and low-resolution cubes were rotated, using the central position
and PA, such that the major axis of the galaxy was aligned with the 
horizontal axis of the image.
We created zeroth moment maps for both rotated cubes, using two times the
measured background noise $\sigma$ for clipping. These cubes are shown 
in the bottom-left panels of Figures 3-10. The
high-resolution zeroth-moment map was also used to create contour levels on top
of the Digital Sky Survey (DSS) J-band image of that region. These are shown in
the top-left panels of Figures 3-10. 

As shown in Section \ref{sec:toymodel}, it is also interesting to compute 
the maximum-brightness-temperature maps of the
galaxies, as they are instrumental in detecting any potential optical
depth issues. The cube was converted from intensities $I$ in mJy/beam to
brightness temperatures $T_B$ in K, using Equation \ref{eqn:JytoK}. 
We created a maximum brightness temperature map by selecting the maximum
temperature along the velocity axis $v$ for each position $(X,z)$. These are 
shown in the top-right panels of Figures 3-10. The middle plot shows the 
maximum
temperature along the above-the-plane height $z$ for each position $(X,v)$. 
This creates an equivalent to a position-velocity diagram, only here it shows 
the maximum brightness-temperature regardless of height $z$. 
The lower panel takes the maximum brightness-temperature along both $v$ and 
$z$, and infers the corresponding optical depth,
based on assumed spin temperature of 150, 200 and 300 Kelvin.
We chose the values to lie close to the mean observed values for the cold 
neutral medium
(CNM) \citep{Dickey2009A}. As such, it forms a diagnostic tool for potential 
self-absorption
issues. We calculated the opacity $\tau_\nu$ using Equation \ref{eqn:tau}, 
which is the inverse of Equation \ref{eqn:Tgood},
\begin{equation}
 \tau_\nu = -\ln\left(1 - T_B / T_\mathrm{spin}\right)\,\,.\label{eqn:tau}
\end{equation}

The integrated position-velocity (XV) diagrams are shown in the bottom-right 
panels of Figures 3-10. These have been created by integrating the rotated
high-resolution cubes along their minor axis and converting them to brightness
temperatures. The right-hand panels show the integrated spectra $S(v)$, which
were created by integrating the XV-diagrams over the major axis. Similarly, we
show the brightness as function of major-axis position in the lower plots, by 
integrating the XV
diagrams over $v$.

The width of the profile at the 20\% and 50\% maximum intensity levels, $W20$
and $W50$, were measured directly from $S(v)$. These were based on the
low-resolution cubes, which were masked using the regions files. The integrated
flux $FI$ was also measured in this cube. We make no use of clipping, in
contrast to \cite{OBrien2010A}, as we wish to include all faint emission.
The total optically thin \hi mass,
$M_\hi$, was derived using
\begin{equation}
 M_\textrm{\hi} = 2.343 \cdot 10^5 \times D^2 \times \textrm{FI} \,\, ,
\end{equation}
where the adopted distance $D$ is in Mpc as shown in Table 
\ref{tbl:opticalproperties} \citep{Draine2011}. 
This value will represent only a lower limit to the true \hi mass.
The derived values for FI, $M_\textrm{\hi}$, $W20$ and $W50$ are shown
in Table \ref{tbl:HISpectrumProperties}.

The systematic velocity v$_\textrm{sys}$ was measured from the XV-diagram in a
method similar to the position finding. For a wide range of possible systemic
velocities v$_\textrm{sys}$, the XV-diagram was rotated $180^\circ$ around that
velocity. The rotated XV-diagram was subtracted from the original, with the
lowest $\chi^2$ was adopted for v$_\textrm{sys}$. The maximum velocity 
v$_\textrm{max}$
was calculated by locating the furthest channel from v$_\textrm{sys}$ that still 
contained flux above two times the $\sigma$ background noise level. The 
results for
v$_\textrm{sys}$ and v$_\textrm{max}$ are also shown in Table 
\ref{tbl:HISpectrumProperties}.
The channel-maps are presented in online Appendix \ref{sec:channelmaps}.

\begin{figure*}
\centering
   \includegraphics[width=0.459\textwidth]{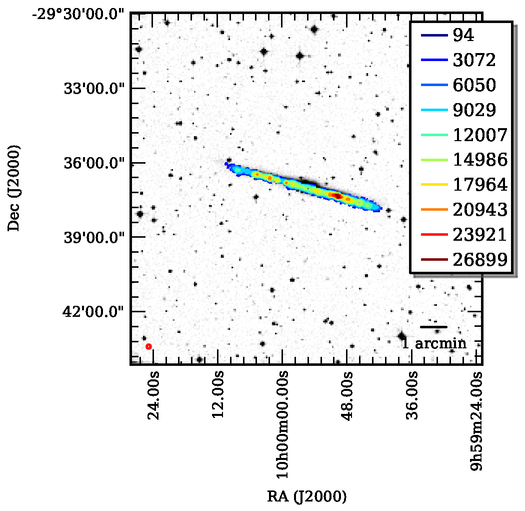}
   \includegraphics[width=0.459\textwidth]{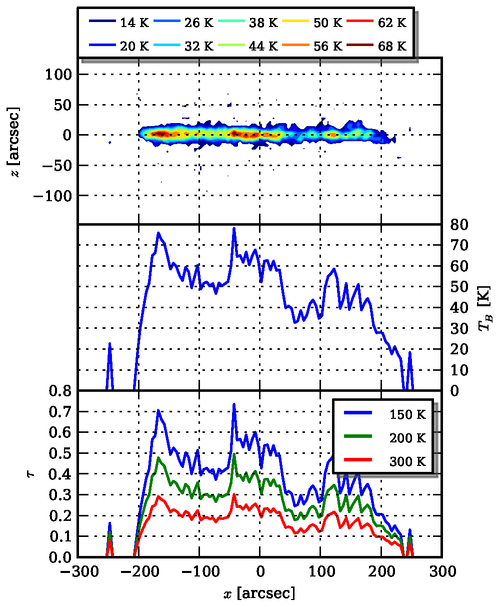}
   \includegraphics[width=0.459\textwidth]{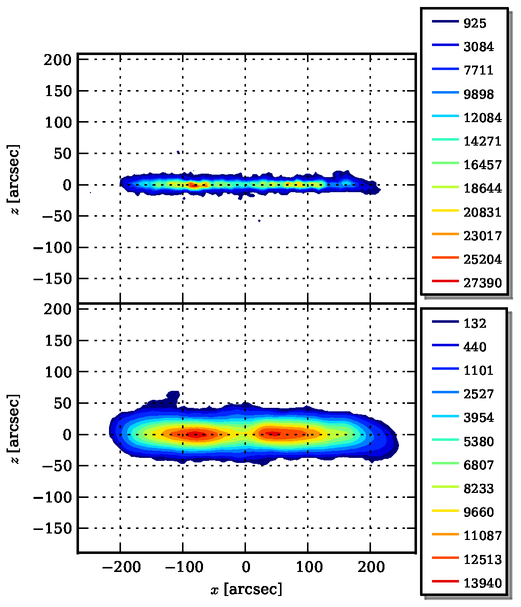}
   \includegraphics[width=0.459\textwidth]{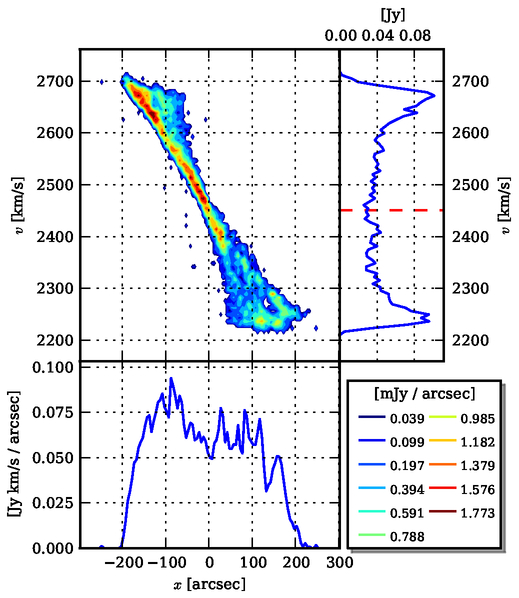} 
\caption[Neutral hydrogen in IC\,2531]{Overview of the neutral hydrogen contents of IC\,2531. Top-left: DSS1 overlaid with contours from the \hi moment 0 map in [K/(km/s)]. Top-right: Top panel shows the maximum temperature along $v$. The middle panel shows the maximum temperature along $z$ with the same scale. The lower plot maximum temperature along $v$ and $z$ and shows the inferred self-absorption for that position, assuming a spin temperature of 125 K. Bottom-left: Top panel shows the moment 0 for the high-resolution cube in [K/(km/s)]. Lower plot shows the moment 0 for the low-resolution cube. Bottom-left: Left panel shows the PV-diagram. Right-top panel shows the integrated flux per velocity. The red dashed line shows v$_\textrm{sys}$. The lower plot shows the integrated flux per position along the major axis.}\label{fig:IC2531}
\end{figure*}

\begin{figure*}
\centering
\includegraphics[width=0.459\textwidth]{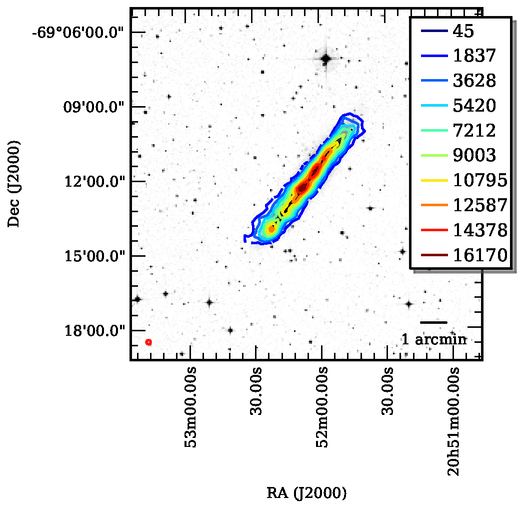}
\includegraphics[width=0.459\textwidth]{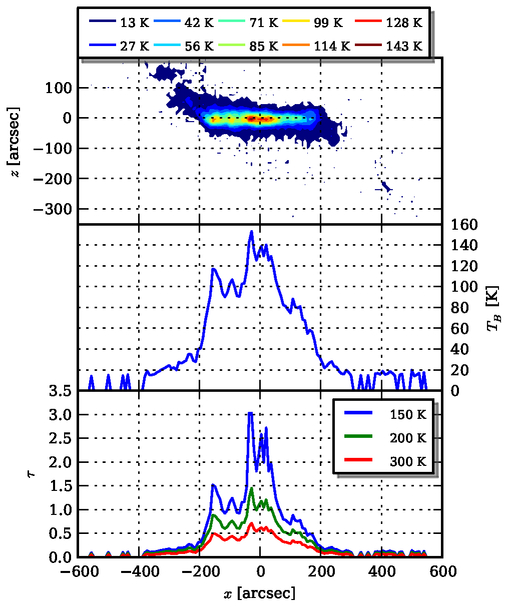}
\includegraphics[width=0.459\textwidth]{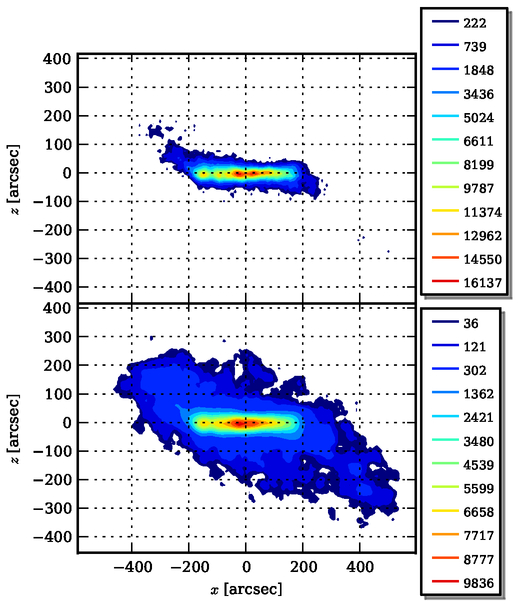}
\includegraphics[width=0.459\textwidth]{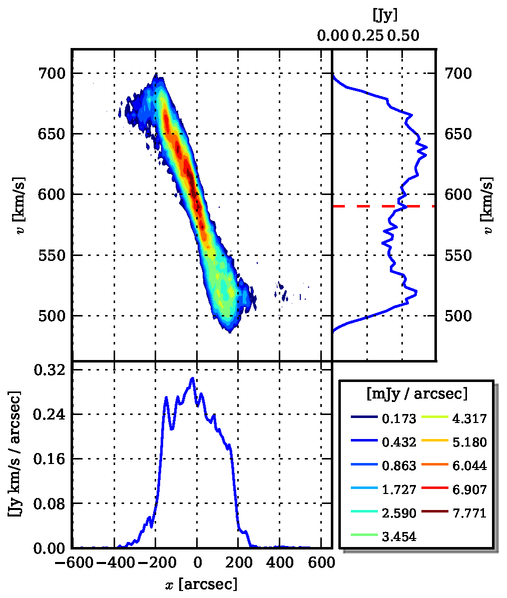}
\caption[Neutral hydrogen in IC\,5052]{Overview of the neutral hydrogen contents of IC\,5052. Top-left: DSS1 overlaid with contours from the \hi moment 0 map in [K/(km/s)]. Top-right: Top panel shows the maximum temperature along $v$. The middle panel shows the maximum temperature along $z$ with the same scale. The lower plot maximum temperature along $v$ and $z$ and shows the inferred self-absorption for that position, assuming a spin temperature of 125 K. Bottom-left: Top panel shows the moment 0 for the high-resolution cube in [K/(km/s)]. Lower plot shows the moment 0 for the low-resolution cube. Bottom-left: Left panel shows the PV-diagram. Right-top panel shows the integrated flux per velocity. The red dashed line shows v$_\textrm{sys}$. The lower plot shows the integrated flux per position along the major axis.}\label{fig:IC5052}
\end{figure*} 

\begin{figure*}
\centering
\includegraphics[width=0.459\textwidth]{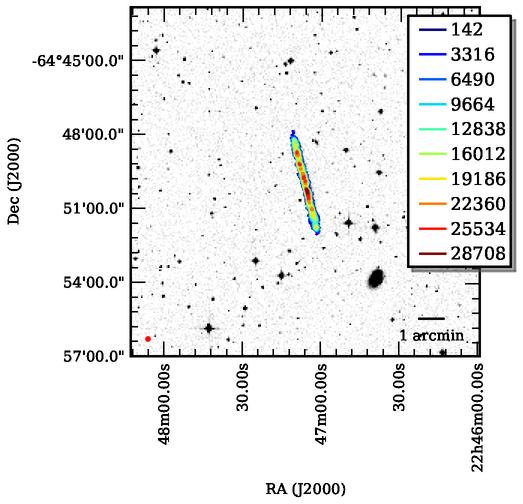}
\includegraphics[width=0.459\textwidth]{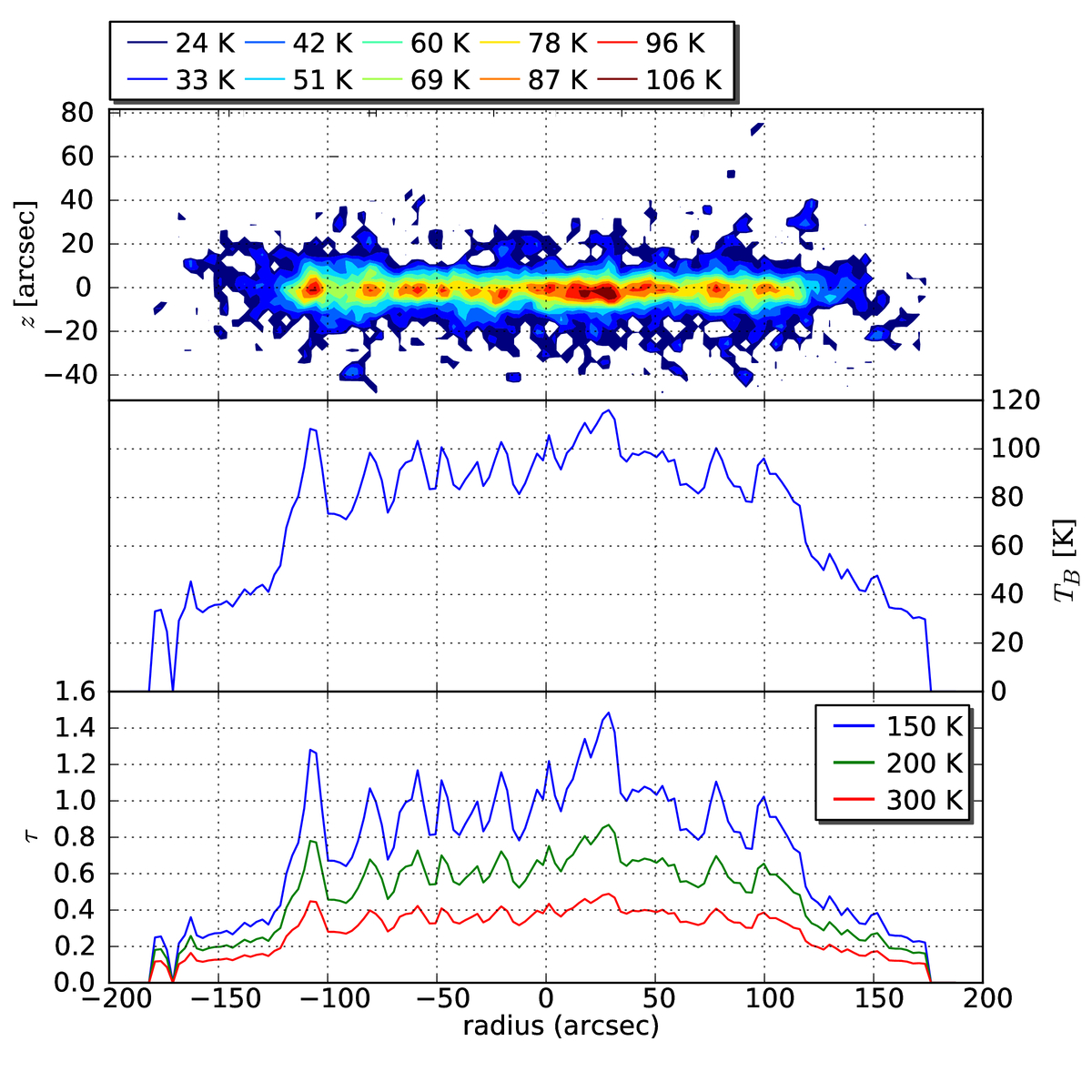}
\includegraphics[width=0.459\textwidth]{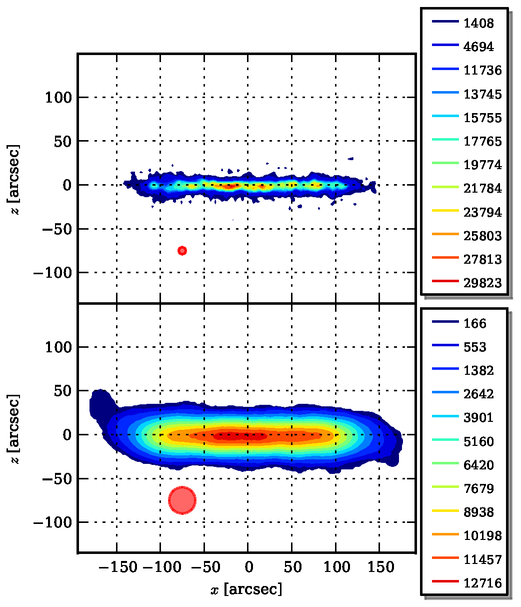}
\includegraphics[width=0.459\textwidth]{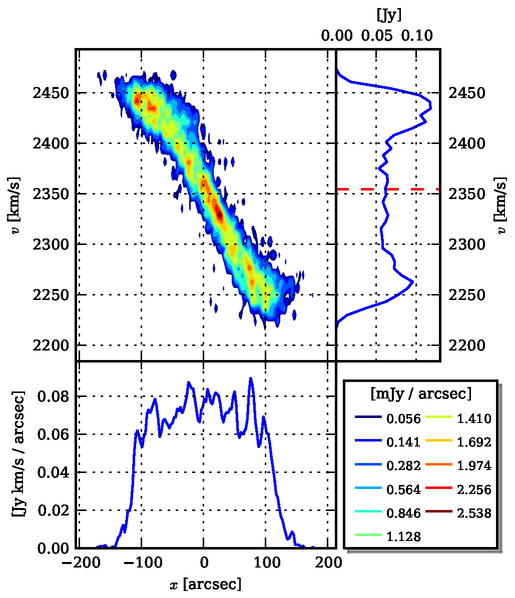}
\caption[Neutral hydrogen in IC\,5249]{Overview of the neutral hydrogen contents of IC\,5249. Top-left: DSS1 overlaid with contours from the \hi moment 0 map in [K/(km/s)]. Top-right: Top panel shows the maximum temperature along $v$. The middle panel shows the maximum temperature along $z$ with the same scale. The lower plot maximum temperature along $v$ and $z$ and shows the inferred self-absorption for that position, assuming a spin temperature of 125 K. Bottom-left: Top panel shows the moment 0 for the high-resolution cube in [K/(km/s)]. Lower plot shows the moment 0 for the low-resolution cube. Bottom-left: Left panel shows the PV-diagram. Right-top panel shows the integrated flux per velocity. The red dashed line shows v$_\textrm{sys}$. The lower plot shows the integrated flux per position along the major axis.}\label{fig:IC5249}
\end{figure*}

\section{Results}\label{sec:HIresults}

\subsection{IC\,2531}\label{sec:HI-IC2531}
Galaxy {IC\,2531} is the only Sc galaxy in our sample. The classification
is different from that of \citet{Bureau1999A} and \citet{OBrien2010A}, where the
galaxy has been classified as a Sb galaxy. 
The classification of Sb occurs first in \citet{Bureau1999A}, who note that 
this classification was not extremely accurate.
It} has the largest
gas mass of the sample, and has a maximum rotation velocity v$_\textrm{max}$
which is twice as high as the rest of the sample (Table
\ref{tbl:HISpectrumProperties}). The \hi layer is remarkably flat, with
the disc seemingly having a smaller scale-height than the stellar disc (Figure
\ref{fig:IC2531} top-left). All of this is supports an Sc classification. 
It is true that the bulge of IC\,2531 is not very weak as in many Sc's, 
so that a Sbc classification cannot be ruled out. In any case, we believe the 
evidence to point at a later classification than Sb.

\citet{BAADBF} have shown that at K-band
IC/,2531 shows signatures of a peaunut-shaped or
boxy spheroid with a X-shape structure in the 
unsharp masked image of the bulge, which would be associated with a bar 
\citep[see e.g.][for a recent review]{Ath15}.
Based on the double horns in the integrated profile, we would expect the galaxy
to be symmetric (Figure \ref{fig:IC2531} bottom-right). The XV-diagram itself is
however asymmetric. Interestingly, HIPASS also reports a more
asymmetric profile. The asymmetry is also present in the analysis of the galaxy
in \citet[their Figure 15]{OBrien2010C}, where the two sides of the derived 
surface density can be seen to deviate strongly.

The XV-diagram shows a so-called ``figure 8''-pattern, for which
\citet{Bureau1997A} suggest that it is due to an extended weak bar.
This bar did not show up in in the H$\alpha$ analysis of
\citet{Bureau1999A}, although they suffered from background problems for this
galaxy. This lack of detection was not fully unexpected, as 
\citet{Bureau1997A} already noted that
the figure extended over a very long range. Bureau \& Freeman speculated that 
the figure
might also be due to a warp, density ring or a spiral arm. \citet{Ann2006A} 
confirmed the presence of a warp in this galaxy. 
We also think it more likely that this
pattern is associated with the warp and does not represent a bar. The warp is 
visible in Figure \ref{fig:IC2531} bottom-left. The K-band observations
mentioned above do show evidence for a central bar, but this this bar does not
appear to extend over much more than the central hole in the \hi distribution.

The XV-diagram is also asymmetric in the rotation curve. The receding side shows
a slowly increasing rotation curve, while the approaching side show a sharp
increase followed by a flat rotation.

The maximum surface brightness temperature map (Figure 
\ref{fig:IC2531} top-right) shows a strong central 
peak and some additional peaks. 
There does not appear to be symmetry in the distribution of these peaks.

\subsection{IC\,5052}\label{sec:HI-IC5052}
{IC\,5052} is a nearby SBd galaxy. The bulge is clearly visible in Figure
\ref{fig:IC5052} top-left. 
We find a previous unreported, line-of-sight warp in the
outskirts of the galaxy. This is particularly clear in the zeroth-moment map of
the low resolution cube (Figure \ref{fig:IC5052} bottom-left). Further
modelling will be required to confirm this as a true warp, although this is 
beyond the scope of this project. 
For now, we note that the channel maps, seen in Figures 
\ref{figure:IC5052channelsa}, \ref{figure:IC5052channelsb}, 
\ref{figure:IC5052channelsc} in online Appendix B, are very similar 
to the fitted model of the line-of-sight warp
in {ESO\,123-G13} \citep{Gentile2003A}. The warp appears to be symmetric
in pitch angle on either side of the galaxy, although it is more extended 
towards the lower-left side.
We have checked various optical and infrared images, but we can find no 
optical counterpart to the warp.

The XV-diagram of this galaxy (Figure \ref{fig:IC5052} bottom-right) 
shows signs of this
warp beyond 250 arcseconds. The main part of the profile looks like a solid-body
rotator. The high- and low velocity sides are asymmetric, which is also clear
from the integrated profile. The shape of the spectrum does agrees well with the
profile from HIPASS.

The maximum surface brightness temperature profile of Figure
\ref{fig:IC5052} top-left shows a plateau around 80-100 K for most of the
inner part of the galaxy. Only the central part shows a higher temperature,
rising to a maximum of 140\,K. This central over-density is also visible in the
zeroth-moment maps (\ref{fig:IC5052} bottom-left).
\citet{OBrien2010A} identify this as a star-forming region.

\begin{figure*}
\centering
\includegraphics[width=0.459\textwidth]{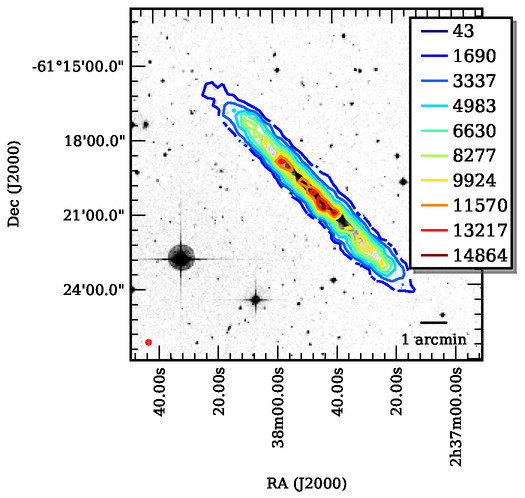}
\includegraphics[width=0.459\textwidth]{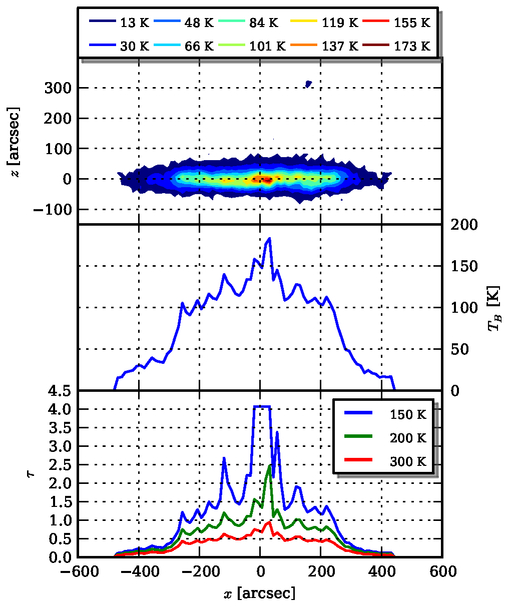}
\includegraphics[width=0.459\textwidth]{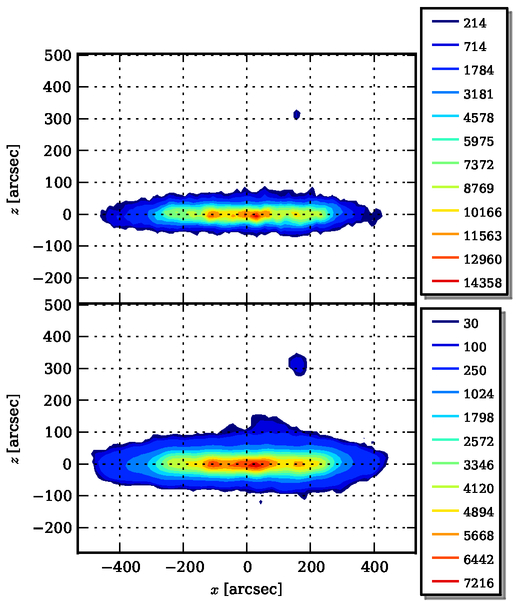}
\includegraphics[width=0.459\textwidth]{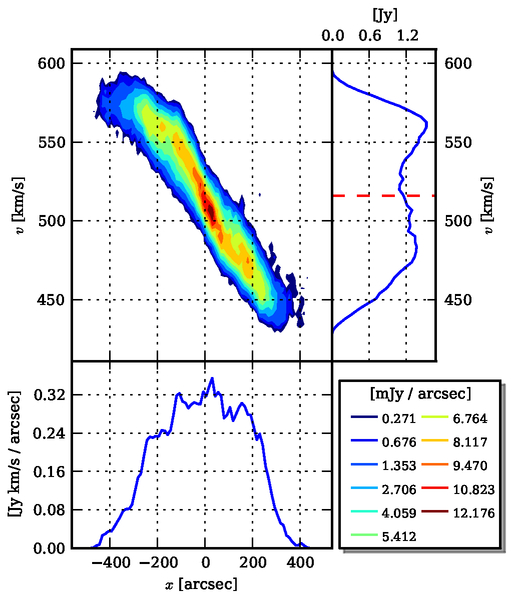}

\caption[Neutral hydrogen in ESO\,115-G021]{Overview of the neutral hydrogen contents of ESO\,115-G021. Top-left: DSS1 overlaid with contours from the \hi moment 0 map in [K/(km/s)]. Top-right: Top panel shows the maximum temperature along $v$. The middle panel shows the maximum temperature along $z$ with the same scale. The lower plot maximum temperature along $v$ and $z$ and shows the inferred self-absorption for that position, assuming a spin temperature of 125 K. Bottom-left: Top panel shows the moment 0 for the high-resolution cube in [K/(km/s)]. Lower plot shows the moment 0 for the low-resolution cube. Bottom-left: Left panel shows the PV-diagram. Right-top panel shows the integrated flux per velocity. The red dashed line shows v$_\textrm{sys}$. The lower plot shows the integrated flux per position along the major axis.}\label{fig:ESO115-G021}
\end{figure*}

\begin{figure*}
\centering
\includegraphics[width=0.459\textwidth]{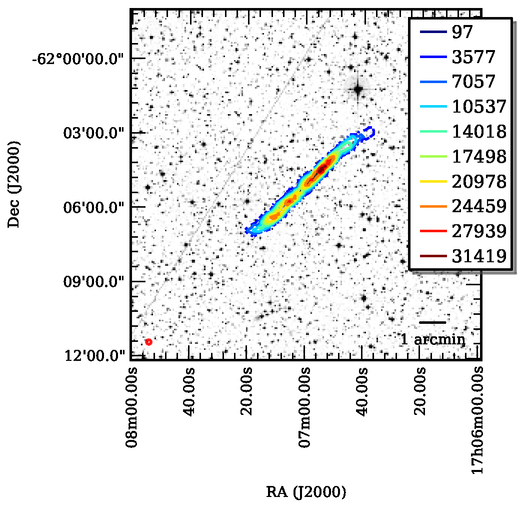}
\includegraphics[width=0.459\textwidth]{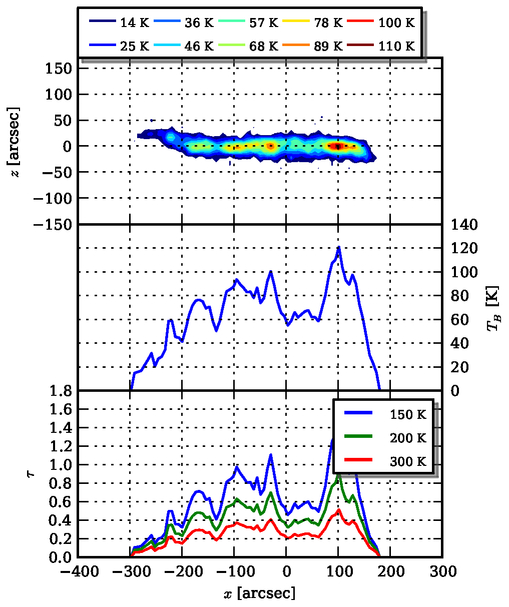}
\includegraphics[width=0.459\textwidth]{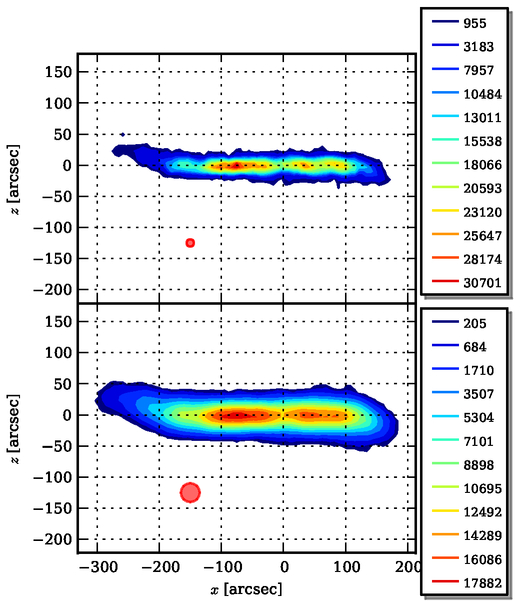}
\includegraphics[width=0.459\textwidth]{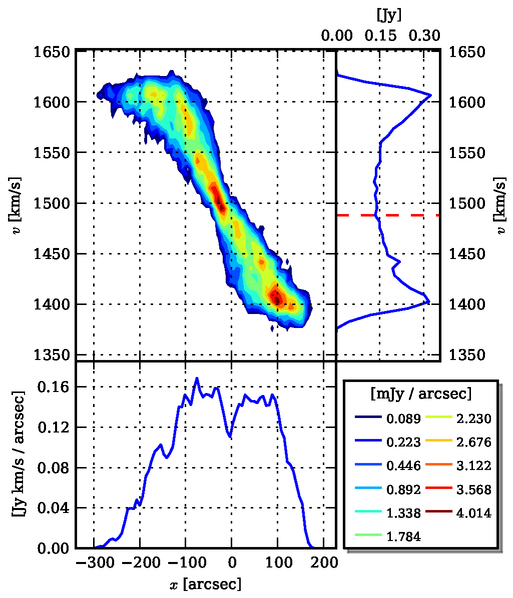}
\caption[Neutral hydrogen in ESO\,138-G014]{Overview of the neutral hydrogen contents of ESO\,138-G014. Top-left: DSS1 overlaid with contours from the \hi moment 0 map in [K/(km/s)]. Top-right: Top panel shows the maximum temperature along $v$. The middle panel shows the maximum temperature along $z$ with the same scale. The lower plot maximum temperature along $v$ and $z$ and shows the inferred self-absorption for that position, assuming a spin temperature of 125 K. Bottom-left: Top panel shows the moment 0 for the high-resolution cube in [K/(km/s)]. Lower plot shows the moment 0 for the low-resolution cube. Bottom-left: Left panel shows the PV-diagram. Right-top panel shows the integrated flux per velocity. The red dashed line shows v$_\textrm{sys}$. The lower plot shows the integrated flux per position along the major axis.}\label{fig:ESO138-G014}
\end{figure*}

\begin{figure*}
\centering
\includegraphics[width=0.459\textwidth]{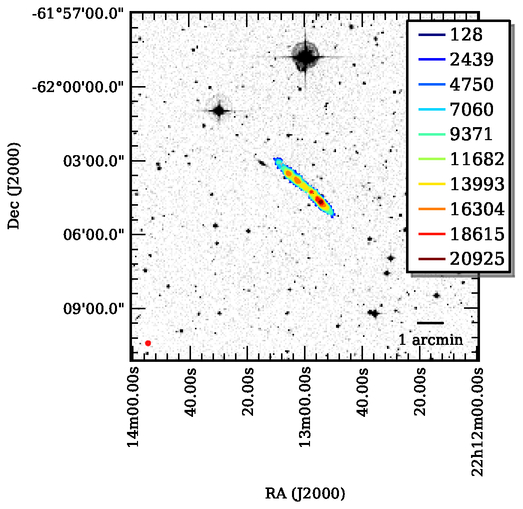}
\includegraphics[width=0.459\textwidth]{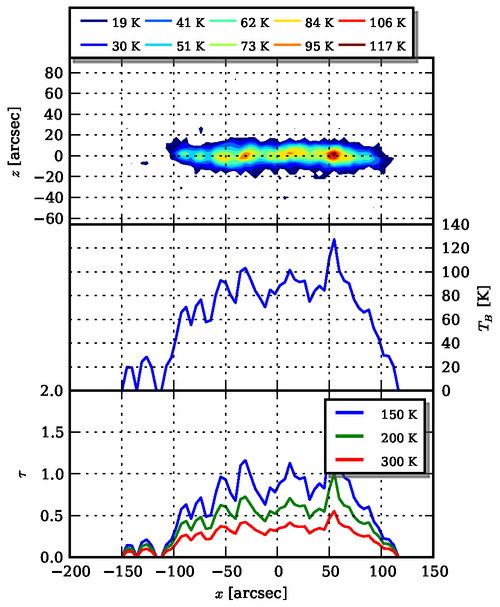}
\includegraphics[width=0.459\textwidth]{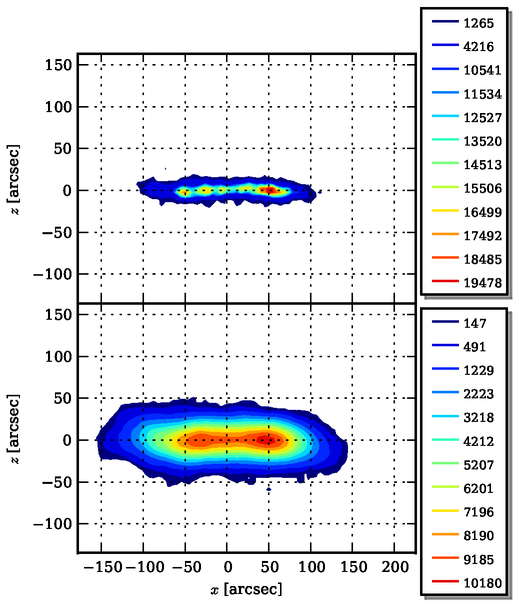}
\includegraphics[width=0.459\textwidth]{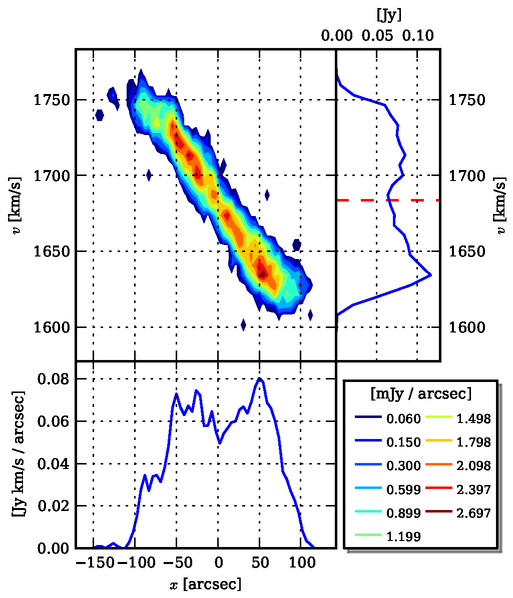}
\caption[Neutral hydrogen in ESO\,146-G014]{Overview of the neutral hydrogen contents of ESO\,146-G014. Top-left: DSS1 overlaid with contours from the \hi moment 0 map in [K/(km/s)]. Top-right: Top panel shows the maximum temperature along $v$. The middle panel shows the maximum temperature along $z$ with the same scale. The lower plot maximum temperature along $v$ and $z$ and shows the inferred self-absorption for that position, assuming a spin temperature of 125 K. Bottom-left: Top panel shows the moment 0 for the high-resolution cube in [K/(km/s)]. Lower plot shows the moment 0 for the low-resolution cube. Bottom-left: Left panel shows the PV-diagram. Right-top panel shows the integrated flux per velocity. The red dashed line shows v$_\textrm{sys}$. The lower plot shows the integrated flux per position along the major axis.}\label{fig:ESO146-G014}
\end{figure*} 

\begin{figure*}
\centering
\includegraphics[width=0.459\textwidth]{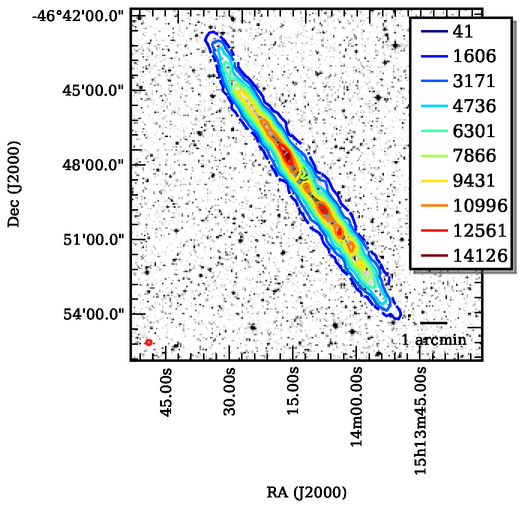}
\includegraphics[width=0.459\textwidth]{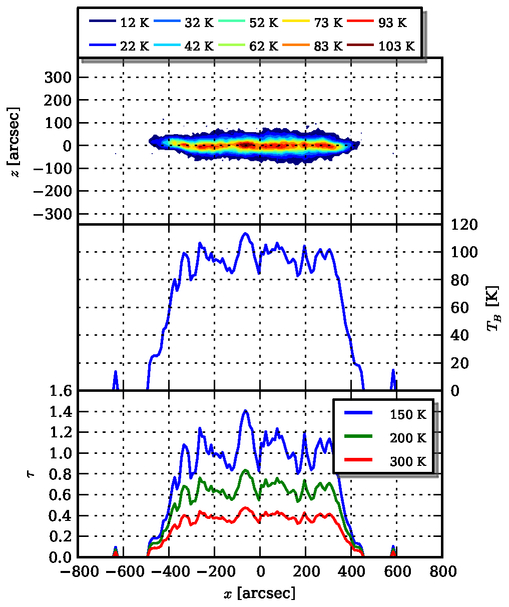}
\includegraphics[width=0.459\textwidth]{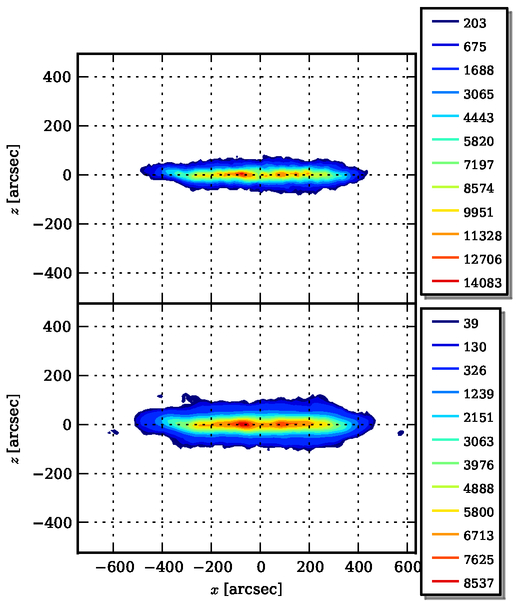}
\includegraphics[width=0.459\textwidth]{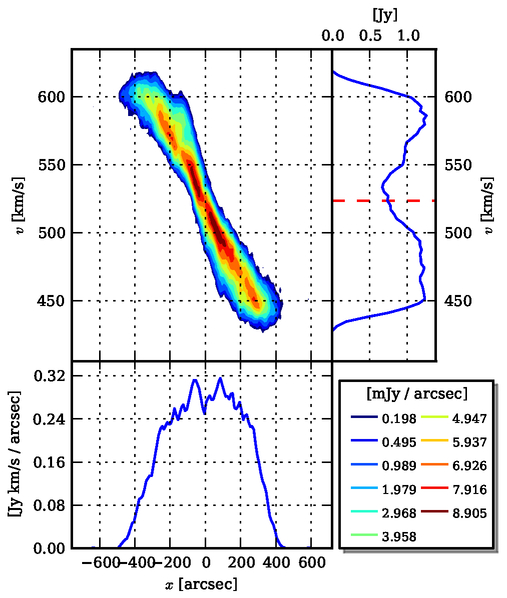}
\caption[Neutral hydrogen in ESO\,274-G001]{Overview of the neutral hydrogen contents of ESO\,274-G001. Top-left: DSS1 overlaid with contours from the \hi moment 0 map in [K/(km/s)]. Top-right: Top panel shows the maximum temperature along $v$. The middle panel shows the maximum temperature along $z$ with the same scale. The lower plot maximum temperature along $v$ and $z$ and shows the inferred self-absorption for that position, assuming a spin temperature of 125 K. Bottom-left: Top panel shows the moment 0 for the high-resolution cube in [K/(km/s)]. Lower plot shows the moment 0 for the low-resolution cube. Bottom-left: Left panel shows the PV-diagram. Right-top panel shows the integrated flux per velocity. The red dashed line shows v$_\textrm{sys}$. The lower plot shows the integrated flux per position along the major axis.}\label{fig:ESO274-G001}
\end{figure*}

\begin{figure*}
\centering
\includegraphics[width=0.459\textwidth]{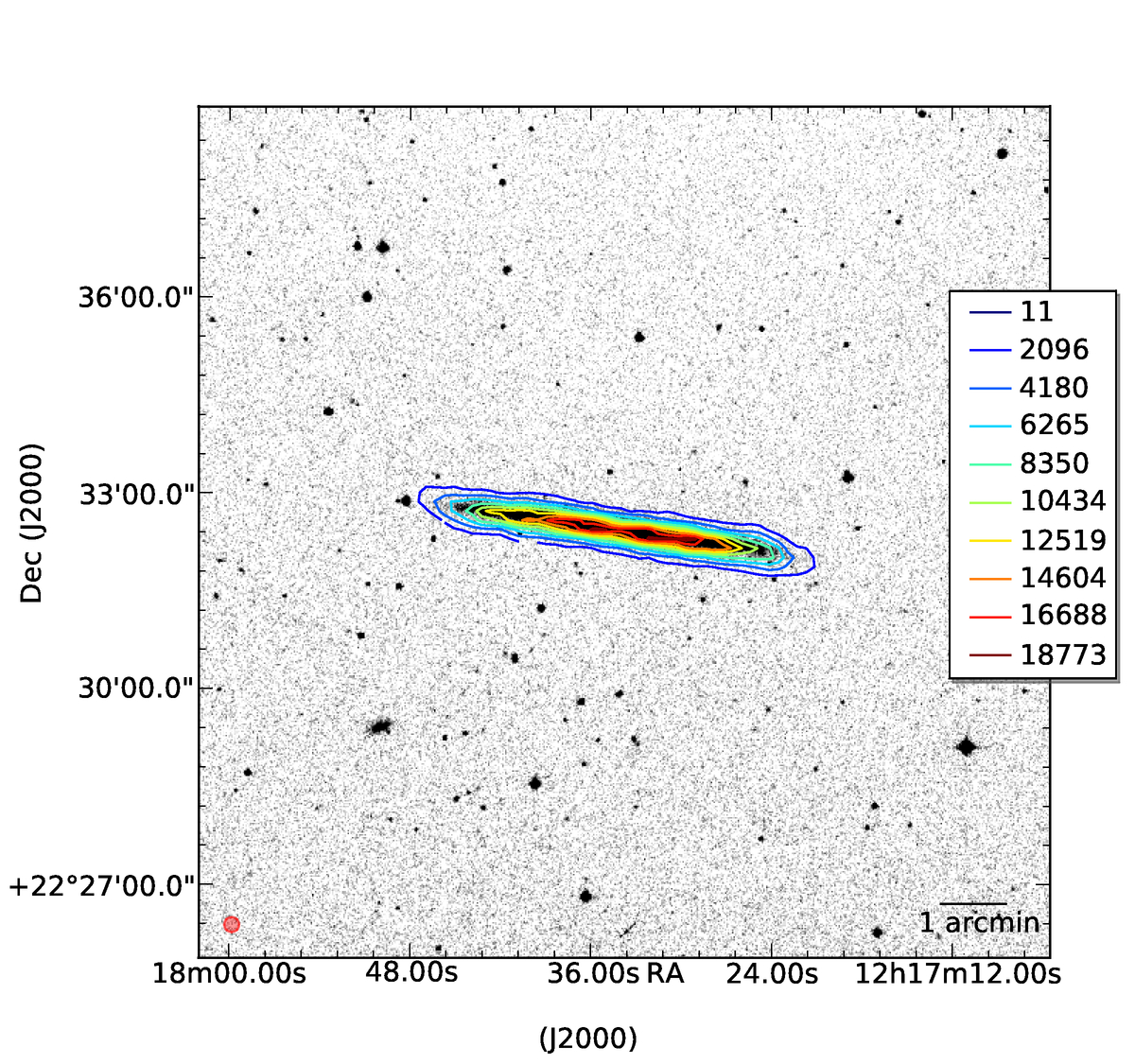}
\includegraphics[width=0.459\textwidth]{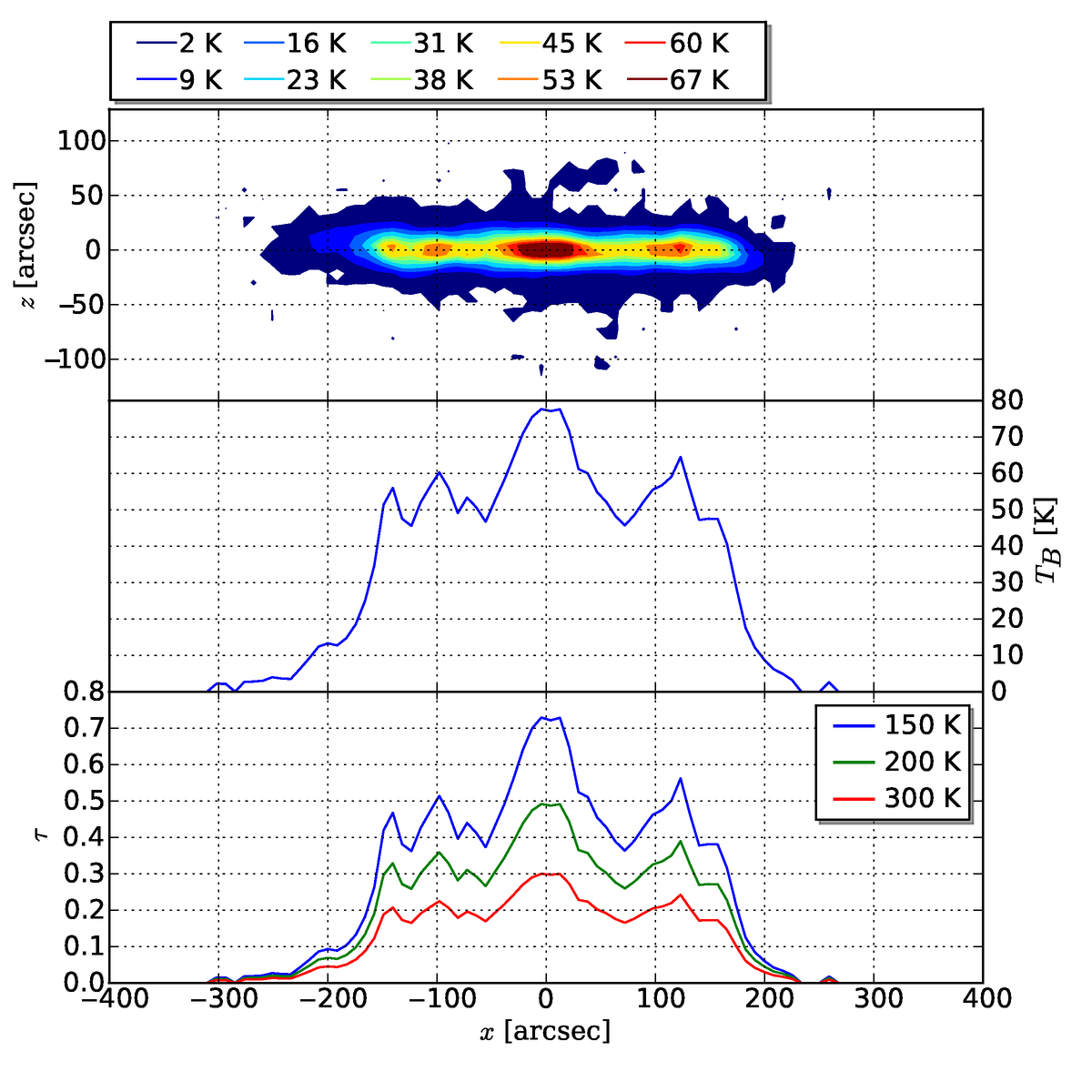}
\includegraphics[width=0.459\textwidth]{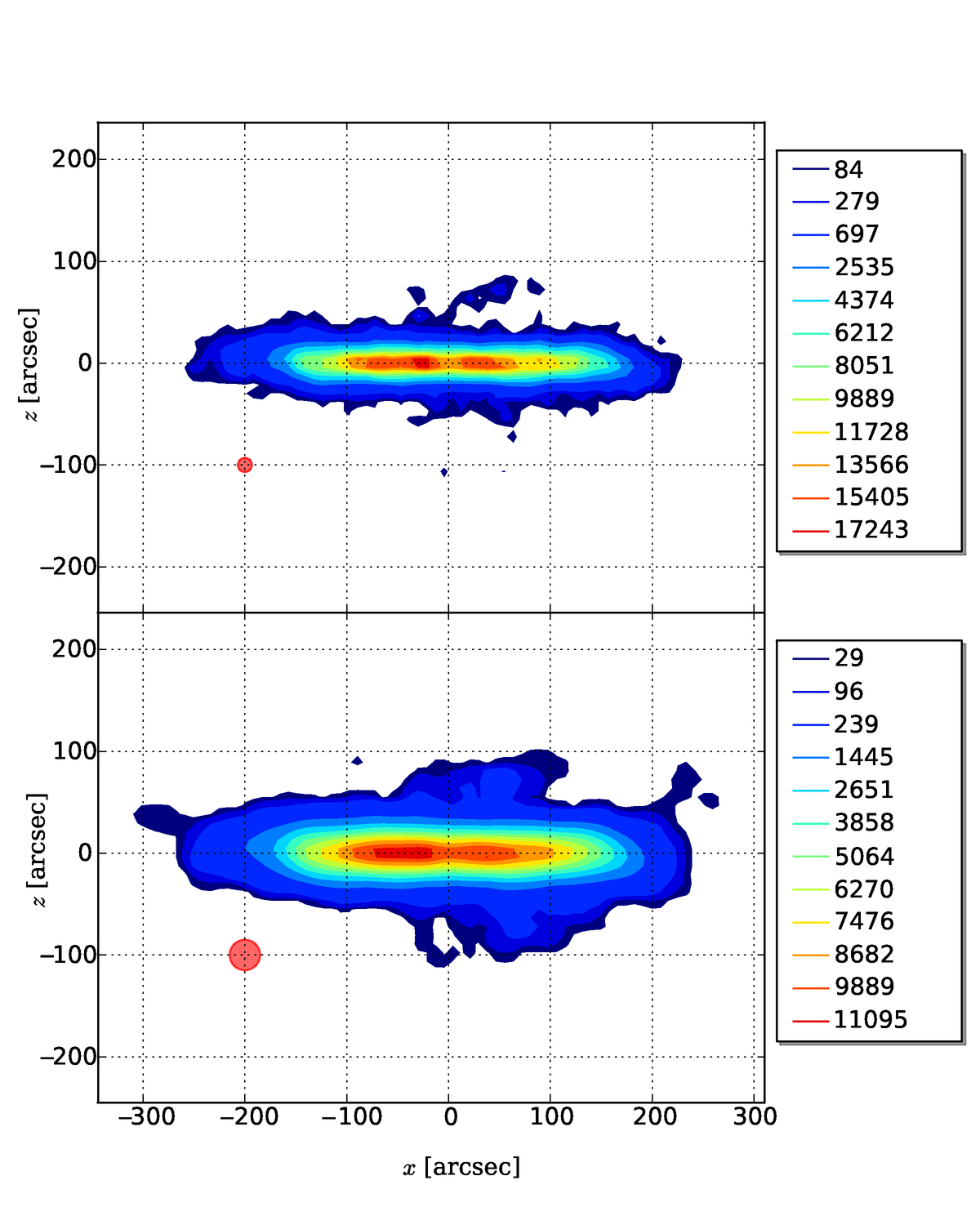}
\includegraphics[width=0.459\textwidth]{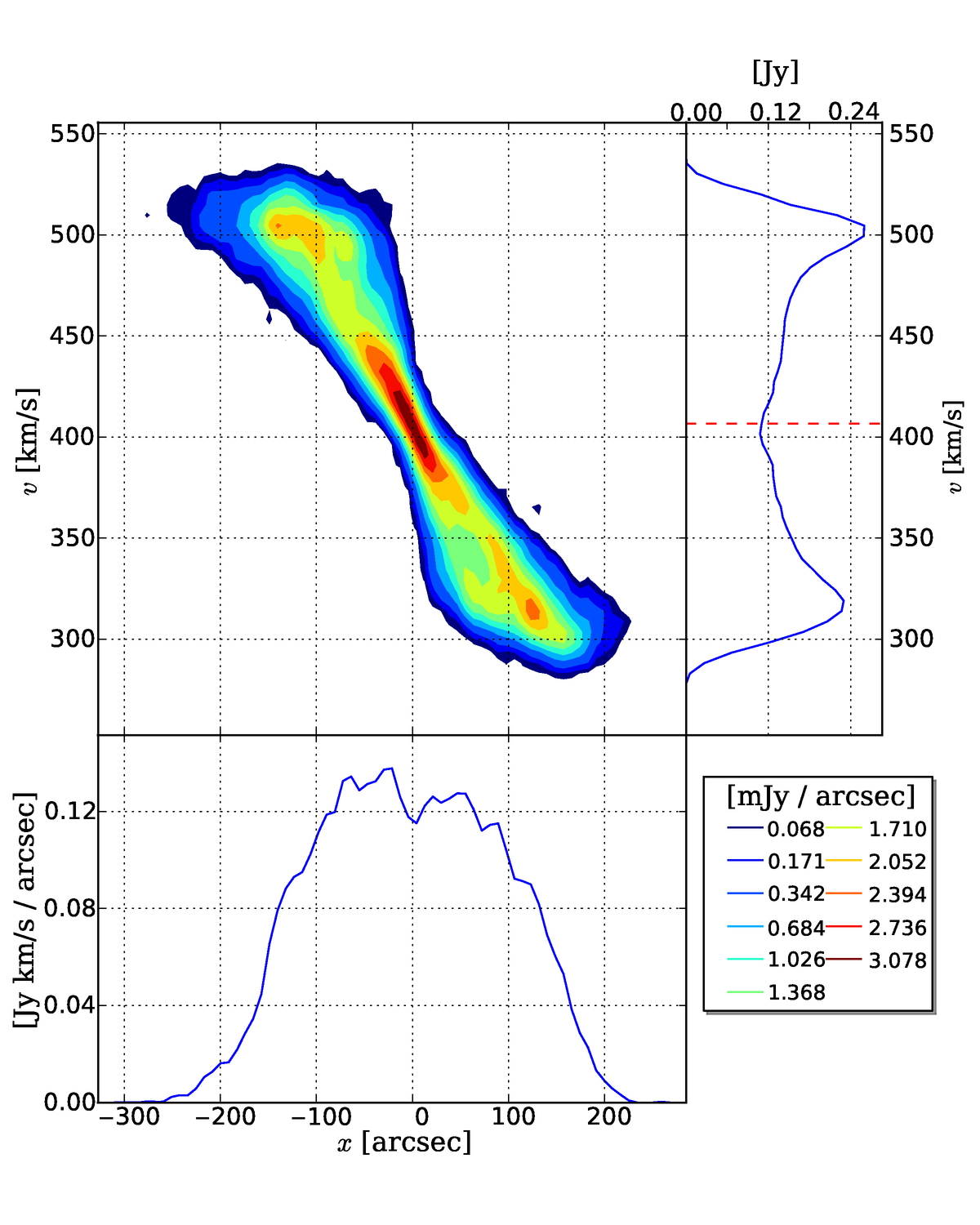}
\caption[Neutral hydrogen in UGC\,7321]{Overview of the neutral hydrogen contents of UGC\,7321. Top-left: DSS1 overlaid with contours from the \hi moment 0 map in [K/(km/s)]. Top-right: Top panel shows the maximum temperature along $v$. The middle panel shows the maximum temperature along $z$ with the same scale. The lower plot maximum temperature along $v$ and $z$ and shows the inferred self-absorption for that position, assuming a spin temperature of 125 K. Bottom-left: Top panel shows the moment 0 for the high-resolution cube in [K/(km/s)]. Lower plot shows the moment 0 for the low-resolution cube. Bottom-left: Left panel shows the PV-diagram. Right-top panel shows the integrated flux per velocity. The red dashed line shows v$_\textrm{sys}$. The lower plot shows the integrated flux per position along the major axis.}\label{fig:UGC7321}
\end{figure*}

\subsection{IC\,5249}
{IC\,5249} is the most distant galaxy in our sample, at a distance of 32\,Mpc. 
The galaxy is only five arcminutes wide.
The maximum S/N of the cube has not been improved since \citet{OBrien2010A}, 
and is
still around ten.

The maximum temperatures exhibit a plateau between 80 and 100\,K, over the 
radius
-100 and 100 arcseconds, along most of the galaxy (Figure
\ref{fig:IC5249} top-right). Another plateau is visible in the 
integrated flux
along the major axis (\ref{fig:IC5249} bottom-right), wherein over that same region the
intensity is seen to hover around 0.06\,Jy km/s/arcsec. The XV-diagram itself
is slightly asymmetric, with the higher velocity side having more \hi gas 
than the
low velocity side.

The rotation curve has been previously analysed by \citet{vanderKruit2001A} 
using \hi and \halp. They find the galaxy to exhibit
a steep rising rotation curve, after which at 1.0 arcmin it flattens out at
$v_\textrm{max} \approx 100$\,km/s.

There is a faint indication of a warp, visible in the lower plot in Figure 
\ref{fig:IC5249} bottom-left.

\subsection{ESO\,115-G021}
{ESO\,115-G021} is a nearby SBdm galaxy at a distance of only 4.9\,Mpc. A lot of
data was available, totaling to 83.5 hours of observation time. Due to this, 
it is a very well
resolved galaxy. The plane of the galaxy is resolved at over $80$ beams and the
galaxy covers $10$ beams in height (see Figure \ref{fig:ESO115-G021} top-left).

The derived spectrum is more symmetric than the HIPASS spectrum, which has a
larger peak at the high-velocity side. There is an external gas cloud at 
02:36:59 -61:18:41
visible in multiple channels. It  shows in both the high and low resolution
cubes (Figure \ref{fig:ESO115-G021} bottom-left) at position (180'', 300'').
In the XV-diagram (Figure \ref{fig:ESO115-G021} bottom-right) 
it is visible as the small
upward peak between 500 and 515\,km/s. This close proximity to the galaxy in 
both
velocity and distance indicates the gas is associated with the galaxy.
The gas cloud has an \hi mass of $3.75 \times
10^5$M$_\odot$.
The cloud is invisible in the DSS, I and V bands (see Paper IV).
The DSS image goes down to a limiting magnitude of 22.5\,R-mag/arcsec$^2$, 
which corresponds to an apparent magnitude of 15.5\,R-mag for the area of 
$\sim600$ arcsec$^2$ that the cloud covers.
This is equal to a solar luminosity of 
$\sim2.84 \times 10^4$\,L$_{\textrm{R},\odot}$.
The cloud thus has a mass-to-light ratio $M_\hi / L_\textrm{R} \geq 13.3$.

The XV-diagram (Figure \ref{fig:ESO115-G021} bottom-right) shows an asymmetry, with the
high velocity side being more extended than the low velocity side. This is most
likely due to the warp, which is visible in the zeroth-moment maps (Figure
\ref{fig:ESO115-G021} bottom-left). The main component of the XV-diagram shows
solid body rotation.

The maximum temperature map (Figure \ref{fig:ESO115-G021} bottom-right) shows
that this warp has a surface brightness of 20-30\,K. The more stable inner parts
of the disc show a maximum temperature plateau around 80-90\,K, with the strong
exception of the region between 0'' and +50'', where there is a rise to a
maximum of 140 K. This region corresponds to a bright region of the DSS image.

\subsection{ESO\,138-G014}
Galaxy {ESO\,138-G014} has the least observing time of all the galaxies in 
our sample.
Nevertheless, the galaxy is well resolved by the telescope. The minor axis
of the galaxy is covered by roughly six beams.

The XV-diagram (\ref{fig:ESO138-G014} bottom-right) is asymmetric, with the receding side
showing an extended component. This is associated with a warp, as visible in 
Figure \ref{fig:ESO138-G014} bottom-left. No optical counterpart to this 
warp is
known \citep{Sanchez2003A}. 

The maximum brightness temperature map forms a plateau throughout most of the
galaxy, averaging between 80 and 100\,K (Figure
\ref{fig:ESO138-G014} bottom-right). The only exception is at the left side,
where a peak of 140\,K is attained. This corresponds to the bright region in the
south-west region in the DSS image (Figure \ref{fig:ESO138-G014} top-left).

\subsection{ESO\,146-G014}\label{sec:HI-ESO146}
Galaxy {ESO\,146-G014} is the faintest galaxy in our sample.
The moment-zero maps (Figure \ref{fig:ESO146-G014} bottom-right)
show that the galaxy does not appear to have a fixed plane. Instead it looks
slightly bent. This is most likely due a strong warp, which can also be seen
in the visible in the DSS J-band image (Figure \ref{fig:ESO146-G014} top-left). 
 
The XV-diagram (Figure \ref{fig:ESO146-G014} top-right) looks symmetrical, although the
integrated spectrum is more asymmetric than the HIPASS spectrum, with a stronger
peak at the low-velocity side. Overall, the galaxy appears to exhibit solid-body
rotation.

The maximum surface brightness temperatures appear to form a plateau over the 
major part of the galaxy, averaging around 90\,K (Figure 
\ref{fig:ESO146-G014} bottom-right). 
On the left-hand side of the galaxy, this maximum drops relatively 
slowly towards the warp.

\subsection{ESO\,274-G001}\label{sec:HI-ESO274}
{ESO\,274-G001} is a SAd galaxy. 
At only 3.02\,Mpc, it is the closest galaxy in our sample. 
It is also has the lowest \hi mass of the galaxies in our sample (Table 
\ref{tbl:HISpectrumProperties}).
Due to its proximity, it is still the largest galaxy in terms of angular 
size, spanning 13 arcminutes in width. 
In height, it is also very well resolved, covering at least 80 arcseconds.

The central hole in the zeroth-moment image (Figure \ref{fig:ESO274-G001} 
top-left) 
is an artifact, due to the removal of a continuum radio source at this 
position. It has a flux density of 48.65 mJy and coincides with (or at 
least is very close) to the symmetry center of both the optical image, the 
\hi distribution and the positon-velocity diagram. The Parkes single-dish 
6-cm flux density is about 50 mJy \citep{WGBE94}, so the source appear to 
be non-thermal.
The most likely explanation is an active galactic nucleus (AGN). 
The incidence of AGN in low luminosity galaxies is probably low; \citet{CCB02}
indicate that the AGN phenomenon is limited to large to E and S0 
galaxies and star powered radio emission to star-forming galaxies. 
AGN activity in low-luminosity 
galaxies seems low, although examples exist suc{\bf h} as NGC\,4395\footnote{We 
thank the referee to pointing out this system.} (see the NASA/IPAC 
Extragalactic Database NED).
It is likely at the galactic centre (see Table \ref{tbl:opticalproperties}) 
and we have therefore manually centred the galaxy on this position.
Any other reasonable choice of the centre will require only a small 
shift, so that our results do not depend critically on this assumption.

AGN are known to exert strong influence on the ISM of the galaxy (e.g. 
\citet{Shulevski2012A} and references therein).
It is therefore interesting to note how regular the \hi disc is, both in the 
zeroth-moment as in maximum-temperature maps.
Most of the inner disc has a maximum surface brightness temperature that forms 
a remarkably flat plateau between 80 and 100\,K (Figure 
\ref{fig:ESO274-G001} bottom-right).
The galaxy is barely visible in most optical and infrared bands, 
indicating the presence of only little stellar mass. 
This may explain why the gas is relatively undisturbed, in comparison 
with our other galaxies.
The galaxy is known to have modest star formation at 
$1.5\times10^{-2}$ M$_\odot$ per year \citep{Cote2009}.

There is a warp visible beyond $\sim280$ arcsec on both sides of the galaxy 
(Figure \ref{fig:ESO274-G001} bottom-left).
The XV-diagram is slightly asymmetric, with more gas at the low velocity side.
This is also visible in the integrated profiles (Figure 
\ref{fig:ESO274-G001} bottom-right).
The shape of the spectrum matches well to HIPASS, although it is more symmetric.

\subsection{UGC\,7321}\label{sec:HI7321}
{UGC\,7321} is a well-known low surface brightness galaxy, which has already been studied using the Westerbork Synthesis Radio Telescope (WSRT) by 
\citet{Garcia-Ruiz2002A} and the VLA by \citet{Uson2003A}. 
The observations by \citet{Uson2003A} form the basis of our work here.
In terms of noise levels, it is the best-resolved galaxy in our sample.
It is also our only northern-sky galaxy.

The zeroth-moment map in Figure \ref{fig:UGC7321} bottom-left shows a 
well-behaved, symmetric galaxy. 
Only at large radii is there some indication of warping, a feature already 
noted by \citet{Matthews2003A}, who also notes it is flaring. 
Additionally, they find evidence that suggests the presence of an \hi halo.
We believe that this halo might be visible as the extended structure on the 
lower figure, between 0 and 110 arcseconds.

The XV-diagram of the galaxy is asymmetric.
The receding side clearly exhibits a flat rotation curve, but  
in contrast the approaching 
side continues to rise over much of the observed extent of the galaxy.

The maximum surface brightness map shows a clear central peak and two 
additional peaks around 100 arcseconds (Figure 
\ref{fig:UGC7321} bottom-right).
These peaks appear to be symmetric around the centre of the galaxy.
The rest of the galaxy forms a plateau, around 60\,K.
We again see that the maximum temperature trails slowly off into the warp on 
the right-hand side.

\section{Discussion}\label{sec:Discussion}
The beam of the high resolution cubes is on average 1-2'' larger
than \citet{OBrien2010A} (Table \ref{tbl:GlobalPropertiesHICubes}). This is due
to most of the new observations being taken in tight baseline configurations. 
There are
therefore more short baselines. The weight in the UV-plane shifts to slightly
smaller scales, resulting in larger beams. The effect is small enough to be of
little importance to this work.
In the high resolution cubes, the noise is significantly lower, thanks to the
additional observations and larger beams. Where no additional observations 
were available, the
noise is (as expected) roughly equal. 

Overall, there is good agreement with \citet{OBrien2010A} on the derived 
central positions, even though we include more faint extended emission in 
our measurements.
The biggest offset is in {ESO\,138-G014} (see Figure 
\ref{fig:ESO138-G014} bottom-left), where the non-symmetric warp is 
affecting the position estimate.
The position angles however do agree well, except that in some cases we have 
added an additional $180^\circ$ because we align the galaxies such that 
the approaching side is aligned to the right-hand side.
Similarly, the systemic velocities v$_\textrm{sys}$ and associated parameters 
$W20$, $W50$ and v$_\textrm{max}$ are all well determined.
Kinematic modelling will have to be performed to find the true centre of each 
galaxy, the derived values here will only serve as first 
estimates for the subsequent analysis in Paper III.

We have also compared the integrated flux $FI$ with the HIPASS catalogue
\citep{Doyle2005A} and the work by \citet{OBrien2010A}. In general our results
match well to HIPASS. {ESO\,274-G001} has the largest difference, with
25 mJy km/s more detected here than in HIPASS. The other major differences are
for {ESO\,115-G021} with a difference of 19\,mJy\,km/s, and {IC
5052} with a difference of 7\,mJy\,km/s. In both cases our observed flux 
again is higher.

These three galaxies are the largest in the sample. The Parkes single dish 
beam has a
angular resolution of 15.5 arcminutes. Most likely, the Parkes beam fails
to cover the entire galaxy, missing the outskirts and is therefore getting a
lower total flux. With that correction in mind, each data cube has successfully
recovered all flux from HIPASS.

\citet{OBrien2010A} report very different values for $FI$. Most striking is
{IC\,5052}, where only 37.9\,mJy\,km/s is reported, against 117.7\,mJy\,km/s 
here. It is 
unclear to us why \citet{OBrien2010A} often reports very different values. Most 
likely, it has to do with the masking of extended emission in their work.
Indeed, our value agrees well with the $\sim 100$\,mJy\,km/s derived by
\citet{Hong13}.

As noted before, the distances to the galaxies are assumed further than in
\citet{OBrien2010A}. This revision is mostly due to new observations, which 
use the tip of the red giant  
branch or the Tully-Fisher relation. This allows us to use cosmology 
independent distances for all galaxies.
That change, plus the higher values for $FI$, combine to have
a large impact on the inferred total mass $M_\textrm{\hi}$ of the galaxy.
Most galaxies now have estimated \hi
masses anywhere between 20 and 300\% higher.

All eight of our galaxies shows signs of warping. 
This is consistent with the work by \citet{Garcia-Ruiz2002A}, 
who found that all galaxies with extended \hi disc have warps.
All of the warps occur at or just beyond the visible stellar disc, 
similar to the findings by \citet{vanderKruit2007A}.
The only possible exception to this rule could be {ESO\,146-G14}, which 
the  warp in the \hi sets in on the sky well
before the end of the stellar disk 
(see Figure \ref{fig:ESO146-G014} top-left).
This is most likely due to the largest amplitude of the warp 
occuring nearer to the line of sight.

\subsection{Self-absorption}
As we pointed out in Section \ref{sec:toymodel}, the 21-cm line of \hi is 
known to be optically thick along many sight lines 
in our own (edge-on) Galaxy \citep{Allen2012A}, and similar results have 
been found in M\,31 and M\,33 \citep{Braun2012A}.
Such self-absorption of the \hi emission not only biases the total hydrogen 
masses, it also raises questions about 
the use of the \hi velocity structure for studies of the gas kinematics.
Nowhere are the answers to such questions more urgent than in the study of 
highly inclined galaxies, where techniques have 
been proposed to ``peel off'' the edges of the observed \hi in order to expose 
the detailed velocity dispersion of the ISM 
over large segments of the galaxy discs in both $R$ and $z$ \citep{OBrien2010B}.

If a moderately inclined galaxy  as M\,31 hides $34\%$ of its \hi mass 
\citep{Braun2012A}, what is the effect on edge-on galaxies?
In the neutral hydrogen, the dominating factor for self-absorption would be 
the spin temperature of the cold neutral medium, which would serve as 
an upper limit to the observable surface brightness temperature.
Consider now the maximum surface brightness maps for {ESO\,274-G001} (Figure 
\ref{fig:ESO274-G001} bottom-right). 
There is a clear plateau around 90 Kelvin throughout almost the entire galaxy. 
At the outskirts the line of sight are much shorter than near the inner parts, 
yet the observed peak temperature remains stable.
A similar behavior can be seen in all other galaxies in our sample.
As a case of even shorter path-lengths, consider \citet[Figure 3]{Allen2012A}, 
where a 'mini-survey' has been performed of \hi, \oh and \co at 30' resolution 
inside the Galaxy.
The observed gas is local, within about 2\,kpc of the sun.
The peak temperature of the \hi is again 90\,K, only this time we are seeing 
the individual clouds. 
Why would such different path-lengths always result in the same brightness 
temperature?

The traditional view is that at these brightness temperatures, column 
densities are reached at which the atomic gas starts to be converted into 
molecular gas, thus providing a natural threshold to the observed temperatures 
\citep{Stecher1967A,Hollenbach1971A,Federman1979A}.
However, a column density is an artificial construct based on the position of 
the observer, and as such, the gas cannot be expected to conform to it.
A volume density threshold would seem to be a much more physically relevant 
quantity.
However, since the path lengths between the local \hi clouds from 
\citet{Allen2012A} and the central parts of the edge-on galaxies 
presented here vary over roughly two orders of magnitude, the volume and 
the inferred density are also likely to vary in the same way.
Any volume density threshold would seem to have difficulty explaining the 
close similarities in the maximum brightness temperatures.

Opacity however can explain this naturally, as we demonstrated in Section 
\ref{sec:toymodel}.
The brightness temperature of 90 Kelvin is in the same range as the median 
spin temperature \citep{Dickey1990A,Dickey2009A,Draine2011}. 
90\,K is then the maximum attainable temperature before the gas turns 
optically thick. 
We are thus seeing optically thick gas in \citet{Allen2012A}.
In Figure \ref{fig:ESO274-G001} bottom-right, the path-lengths have 
become so long compared to the sizes of the clouds, that each line is 
bound to hit an optically thick cloud.
Considering the radii at which these can be seen, a lot more than $30\%$ 
of the galaxy \hi mass could be hidden in and behind these clouds.

The self-absorption will also influence the observed kinematics and structure 
of the disc.
As the brightest parts are of the gas are scaled down by the self-absorption, 
a fit to the thickness of the disc can be deceived into detecting a thicker 
disc than is actually present.
Similarly the velocity dispersion can appear larger than it would have been 
had the gas been optically thin.
Because of these issues, we do not model the galaxies in this paper, but 
will first develop software that can model self-absorption and that will
be the subject of Paper II.

\section*{Acknowledgments}
SPCP is grateful to the Space Telescope Science Institute, Baltimore, USA, the 
Research School for Astronomy and Astrophysics, Australian National University, 
Canberra, Australia, and the Instituto de Astrofisica de Canarias, La Laguna, 
Tenerife, Spain, for hospitality and support during  short and extended
working visits in the course of his PhD thesis research. He thanks
Roelof de Jong and Ron Allen for help and support during an earlier 
period as visiting student at Johns Hopkins University and 
the Physics and Astronomy Department, Krieger School of Arts and Sciences 
for this appointment.

PCK thanks the directors of these same institutions and his local hosts
Ron Allen, Ken Freeman and Johan Knapen for hospitality and support
during many work visits over the years, of which most were 
directly or indirectly related to the research presented in this series op 
papers.

Work visits by SPCP and PCK have been supported by an annual grant 
from the Faculty of Mathematics and Natural Sciences of 
the University of Groningen to PCK accompanying of his distinguished Jacobus 
C. Kapteyn professorhip and by the Leids Kerkhoven-Bosscha Fonds. PCK's work
visits were also supported by an annual grant from the Area  of Exact 
Sciences of the Netherlands Organisation for Scientific Research (NWO) in 
compensation for his membership of its Board.

%\begin{thebibliography}{99}

%\setlength{\bibsep}{0.1em}
\bibliography{refsI}
\bibliographystyle{mn2e}

%\end{thebibliography}

\bsp

\label{lastpage}

\end{document}